\documentclass[aps,prx,twocolumn,superscriptaddress,showpacs,floatfix,longbibliography]{revtex4-2}
\usepackage{amsmath,amssymb,amsfonts,float,graphics,epsfig,epstopdf,color,verbatim,tabularx,bm,multirow,appendix,wasysym}
\usepackage{amsthm}
\usepackage[utf8]{inputenc}
\usepackage{CJKutf8}
\usepackage[T1]{fontenc}
\usepackage{xcolor}
\usepackage{color}

\usepackage{dsfont}
\usepackage{textcomp}
\usepackage{yfonts}
\usepackage{bm}
\usepackage{subfigure}
\usepackage{mathrsfs}
\usepackage{graphicx}
\usepackage{verbatim}
\usepackage{hyperref}
\usepackage{multirow}
\usepackage{braket}
\usepackage[normalem]{ulem}
\usepackage{tikz}
	\usetikzlibrary{calc}
	\usetikzlibrary{shapes.multipart}

\usepackage[most]{tcolorbox}

\newtcolorbox{redframebox}{
  colback=red!10, 
  frame hidden, 
  boxsep=5pt, 
  left=5pt,right=5pt,top=5pt,bottom=5pt, 
  arc=3mm, 
  boxrule=0.5pt, 
  breakable, 
  enhanced jigsaw, 
  sharp corners=downhill, 
}

\newcommand{\comments}[1]{}

\usepackage{natbib} 
\begin{document}

\title{Addressing general measurements in quantum Monte Carlo}

\author{Zhiyan Wang}
\affiliation{Department of Physics, School of Science and Research Center for Industries of the Future, Westlake University, Hangzhou 310030,  China}
\affiliation{State Key Laboratory of Surface Physics and Department of Physics, Fudan University, Shanghai 200438, China}
\affiliation{Institute of Natural Sciences, Westlake Institute for Advanced Study, Hangzhou 310024, China}

\author{Zenan Liu}
\email{liuzenan@westlake.edu.cn}
\affiliation{Department of Physics, School of Science and Research Center for Industries of the Future, Westlake University, Hangzhou 310030,  China}
\affiliation{Institute of Natural Sciences, Westlake Institute for Advanced Study, Hangzhou 310024, China}

\author{Bin-Bin Mao}
\affiliation{School of Foundational Education, University of Health and Rehabilitation Sciences, Qingdao 266000, China}

\author{Zhe Wang}
\affiliation{Department of Physics, School of Science and Research Center for Industries of the Future, Westlake University, Hangzhou 310030,  China}
\affiliation{Institute of Natural Sciences, Westlake Institute for Advanced Study, Hangzhou 310024, China}

\author{Zheng Yan}
\email{zhengyan@westlake.edu.cn}

\affiliation{Department of Physics, School of Science and Research Center for Industries of the Future, Westlake University, Hangzhou 310030,  China}
\affiliation{Institute of Natural Sciences, Westlake Institute for Advanced Study, Hangzhou 310024, China}

\begin{abstract}
Quantum Monte Carlo is one of the most promising approaches for dealing with large-scale quantum many-body systems. It has played an extremely important role in understanding strongly correlated physics. However, two fundamental problems, namely the sign problem and general measurement issues, have seriously hampered its scope of application.
We propose a universal scheme to tackle the problems of general measurement. The target observables are expressed as the ratio of two types of partition functions $\langle \mathrm{O} \rangle=\bar{Z}/Z$, where $\bar{Z}=\mathrm{tr} (\mathrm{Oe^{-\beta H}})$ and $Z=\mathrm{tr} (\mathrm{e^{-\beta H}})$. These two partition functions can be estimated separately within the reweight-annealing frame, and then be connected by an easily solvable reference point. We have successfully applied this scheme to XXZ model and transverse field Ising model, from 1D to 2D systems, from two-body to multi-body correlations and even non-local disorder operators, and from equal-time to imaginary-time correlations. The reweighting path is not limited to physical parameters, but also works for space and time. 
Essentially, this scheme solves the long-standing problem of calculating the overlap between different distribution functions in mathematical statistics, which can be widely used in statistical problems, such as quantum many-body computation, big data and machine learning. \\
\end{abstract}

\date{\today}
\maketitle

\noindent\textbf{Introduction}\\
Quantum Monte Carlo (QMC) is a highly promising numerical method without approximations for large-scale or high-dimensional quantum many-body systems, capable of simulating complex systems with an exponential degree of freedom while maintaining polynomial computation complexity~\cite{ceperley1986quantum,sandvik1991quantum,Ceperley_rmp1995,gubernatis2016quantum,Foulkes_RMP2001,Deng2002Cluster,Sandvik2010Computational,Carlson_rmp2015,DL1_OlavF_2002,Evertz2003loop,Assaad_book2008,Yan2019,ZY2020improved,prokof1998exact,Nicola2022Efficient,Melko2013Stochastic,Blankenbecler1981Monte,Scalapino1981Monte,Hirsch1985,Zhang1999Finite,He2019Finite,Hui2016Quantum,Ma2018Anomalous,cheng2022fractional,Ding2018Engineering}.
Despite the maturity of QMC techniques after decades of development~\cite{Nicola2022Efficient,Ejaaz2024Stochastic,Sandvik2003Stochastic,xu2024loop,fan2023clock,sandvik2019stochastic,Sun2024Delay,sun2024boosting,song2024extended,Li2015Solving,HongYao3,Liu2017Self,Xu2017Self,SHAO2023Progress,Deng2023Improved,Deng2024Diagnosing,chen2024tensor}, there remain two essential challenges that greatly limit the application of QMC. The first is the notorious sign problem \cite{Sugar1990exp,takasu1986monte,hatano1992representation,PhysRevB.92.045110,Sandvik2000sign,pan2022sign,ma2024defining,2005CongjunWu,Wei2016Majorana,Wessel2018Thermodynamic,Emidio2020Reduction,Li2015Solving,HongYao3,Li2015Fermion,Wan2022Mitigating,zhang2022Fermion,Wessel2017,Alet2016sign,mondaini2022quantum}, and the second is the issue of general (off-diagonal) measurements~\cite{Sandvik2010Computational,avella2012strongly,Evertz2003loop,sandvik2019stochastic,Melko2013Stochastic}. 

In this work, we will focus on the enduring challenge of measuring general (off-diagonal) observables. The target is how to extract as more as information from the QMC samplings.
Unlike other numerical methods, QMC cannot directly obtain the wave-function of ground state. Typically, the evaluation of a physical quantity $\langle \mathrm{O} \rangle$ in QMC is derived as follows: $\langle \mathrm{O} \rangle=\mathrm{tr} (\mathrm{Oe^{-\beta H}})/Z$, where $Z=\mathrm{tr} (\mathrm{e^{-\beta H}})$ is the partition function (PF), $\beta$ is the inverse temperature and $H$ is the Hamiltonian. For simplicity, we define $\bar{Z}=\mathrm{tr} (\mathrm{Oe^{-\beta H}})$, hence $\langle \mathrm{O} \rangle=\bar{Z}/Z $. 

In a standard QMC framework, the partition function $Z$ can be generally decomposed into the sum of all the weights, i.e. $Z=\sum_i W_i$. If the operator $\mathrm{O}$ can be treated as a number $\mathrm{O}_i$ under the configuration of $W_i$, which corresponds to a diagonal measurement, the physical quantity can be readily estimated in the form 
\begin{equation}
    \langle \mathrm{O} \rangle=\frac{\bar{Z}}{Z}=\frac{\sum_i \mathrm{O}_i W_i}{Z}  
\end{equation}
In this way, the value $\mathrm{O}_i$ can be directly obtained when we sample the configurations $W_i$ of the PF, making diagonal measurements straightforward in the QMC framework. In the case of diagonal measurement, it is clear that two PFs, $\bar{Z}=\sum_i O_i W_i$ and $Z=\sum_i W_i$, share the same set of configurations $\{W_i\}$, but differ in their associated values, with $O_i$ for $\bar{Z}$ and $1$ for $Z$. Consequently, sampling the configurations $\{W_i\}$ is sufficient to capture the expectation value $\langle \mathrm{O} \rangle=\bar{Z}/Z$.

\begin{figure*}[htp]
\centering
\includegraphics[width=0.9\textwidth]{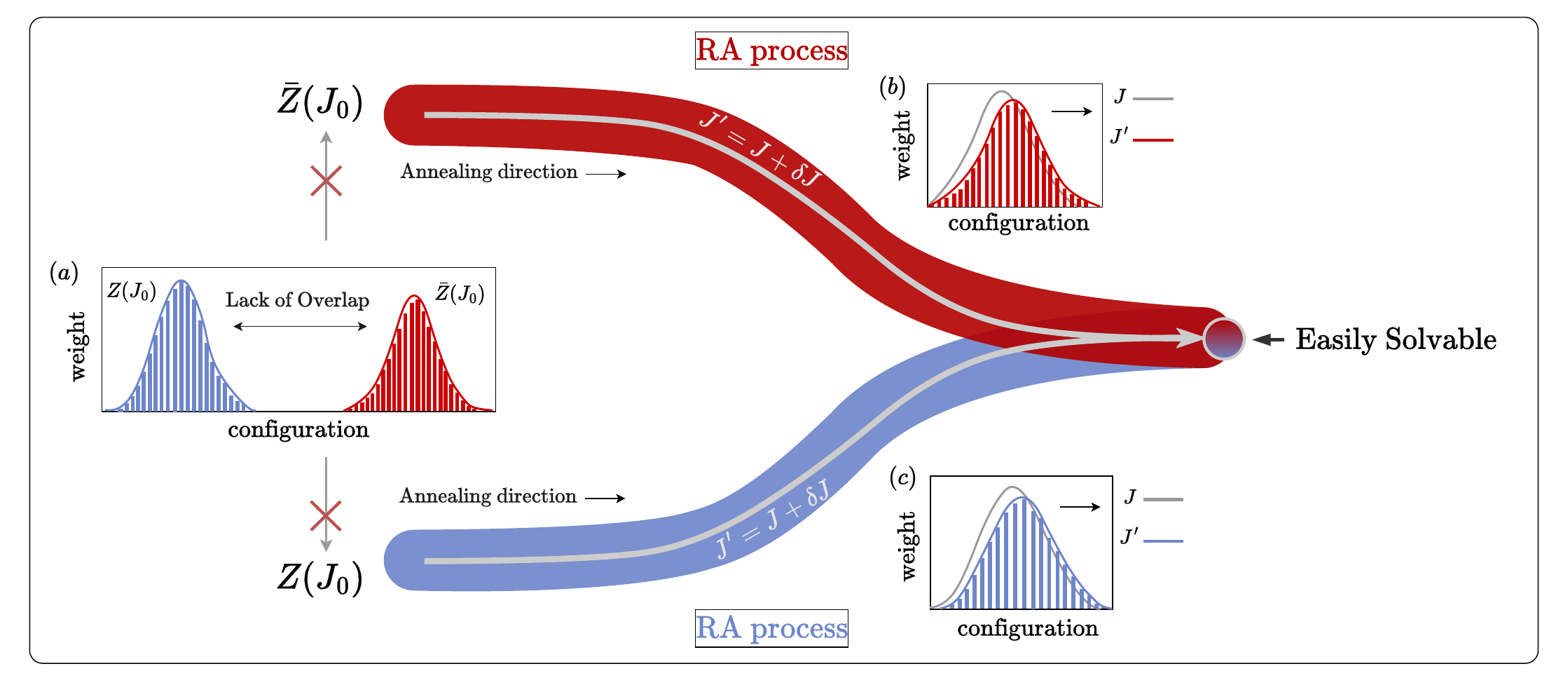}
\caption{Schematic of the bipartite reweight-annealing process. 
Direct evaluation of the ratio $\bar{Z}(J_0)/Z(J_0)$ is generally infeasible, since the corresponding weight distributions at $J_0$ exhibit a lack of overlap, as sketched in (a). Instead, two independent reweight-annealing (RA) processes are performed: the red path corresponds to $\bar{Z}(J)/\bar{Z}(J_0)$, and the blue path corresponds to $Z(J)/Z(J_0)$, as indicated by the colored RA process labels. Along each path, the system is gradually evolved in the annealing direction through small parameter shifts $J’ = J + \delta J$, ensuring sufficient overlap between adjacent distributions, as illustrated in (b) and (c). Once the annealing paths reach an easily solvable reference point, the target ratio $\bar{Z}(J_0)/Z(J_0)$ can be reconstructed.} 
\label{fig:rean}
\end{figure*}

However, the situation would deteriorate significantly during off-diagonal measurements. Off-diagonal operators typically alter the existing configurations $\{ W_i \}$ of $Z=\sum_i W_i$, resulting in new configurations $\{ W'_i \}$ for $\bar{Z}=\sum_i W'_i$ that are entirely distinct from the original set $\{ W_i \}$. This implies that we are unable to obtain samples $\{ W'_i \}$ within the framework of conventional QMC methods,  which are designed to sample from $\{W_i\}$.  As shown in Fig. \ref{fig:rean} (a), two PFs no longer share the same configurations, making it impossible to simulate their ratio directly as in the diagonal case. Furthermore, it is usually impossible to design updates between $\{ W_i \}$ and $\{ W'_i \}$ in QMC algorithms (If you can realize the updates between $\{ W_i \}$ and $\{ W'_i \}$, the ratio $\bar{Z}/Z$ then can be obtained, such as the QMC algorithm for entanglement entropy~\cite{humeniuk2012quantum}). This represents the fundamental challenge in the off-diagonal measurements.

In some special cases, certain off-diagonal observables can be extracted in ingenious ways.  For instance, determinant QMC (DQMC) can efficiently calculate two-body Green’s functions, where higher-order correlators can be derived using Wick’s theorem~\cite{Assaad_book2008, CCChang2013, HongYao3, Sun2024Delay}.  This works particularly well in fermionic systems, where off-diagonal Green’s functions such as $\langle c_i c_j^\dagger\rangle$ are naturally embedded in the sampling weight. 
We now fix the measurement problem in other Monte Carlo methods, such as diagrammatic QMC~\cite{van2010diagrammatic,kozik2010diagrammatic,werner2009diagrammatic}, world-line QMC and stochastic series expansion (SSE) formulations, in which off-diagonal measurements pose a much more severe challenge. 
Despite numerous efforts in the past, two-body Green's functions can be obtained only in certain cases within the framework of worm-like QMC~\cite{prokof2010worm,Boninsegni2006worm,PhysRevB.92.155102,Dorneich2001accessing,zhou2021amplitude,wenjing2021measuring,DL1_OlavF_2002,DL2_OlavF_2003,DL3_Henelius2002,Huang2020Worm}. The reason is that the configurations in the worm-like update process can be treated as samplings of the two-point Green's function. However, multi-body Green's functions remain challenging to be extracted even with this specialized approach and the worm-like algorithm only works in several models.
In addition, if the off-diagonal operator to be measured is a part of the Hamiltonian, it can be estimated through the sampling process~\cite{sandvik1992generalization}. As an instance, $\langle S^x \rangle$ can be measured in a transverse field Ising model (TFIM) ~\cite{yan2023emergent,ZYan2022}.
Another example is that, in the stochastic series expansion (SSE) method, the energy value can be calculated directly by counting the number of operators in the space-time configurations~\cite{sandvik1991quantum,sandvik2019stochastic,Sandvik2010Computational}. 
Although the importance of off-diagonal observables in quantum systems, there is currently no general method for measuring arbitrary operators in QMC, even though a lot of effort has been devoted to it over the past decades.

Recently, a newly proposed method -- reweight-annealing (RA)~\cite{ding2024reweight} has been successfully applied to determine the ratio of two same-type PFs at different parameters. In the reweight-annealing method, as shown in Fig. \ref{fig:rean} (b) and (c), the PF at the parameter $J'$ can be estimated using the value of PF at another parameter $J$ by resetting the weights. 
\begin{equation}
   \frac{Z(J')}{Z(J)}=\bigg\langle \frac{W(J')}{W(J)} \bigg\rangle_{J}
\label{eq:main2}
\end{equation} 
This equation states that the ratio $Z(J')/Z(J)$ is obtained by averaging $W(J')/W(J)$ over configurations sampled from the ensemble at $J$ (i.e.$\langle\cdot\rangle_J$). However, this equation works well only when the distributions $Z(J')$ and $Z(J)$ are adjacent.  
In this context, the importance sampling can be maintained~\cite{neal2001annealed}. Therefore, if the target parameters $J$ and $J'$ are far away from each other, a series of intermediate parameters $\{ J_i \}$ need to be inserted to split the reweighting process by gradually moving from $J$ to $J'$. This can be expressed as 
$Z(J')/Z(J)=Z(J')/Z(J_1)\times Z(J_1)/Z(J_2)\times ... Z(J_i)/Z(J_{i+1}) ... \times Z(J_n)/Z(J)$. Since the entire process involves annealing from one parameter to another with iterative reweighting, it is dubbed as "reweight-annealing" ~\cite{ding2024reweight}. The similar spirit of reweighting also has been developed in the high-energy physics and other fields~\cite{deForcrand2001Hooft,deForcrand2005Measuring,deForcrand2005Precision,Michele2003String,dai2024residual,Mon1985Direct}. Once a reference point $Z(J)$ is known, $Z(J')$ can be calculated through the ratio. It has been proved that the computation complexity of the RA method is polynomial if the ratio of two closest $Z(J)$ and $Z(J')$ is fixed in the division strategies~\cite{ding2024reweight}. More specifically, in the QMC simulation, maintaining the ratio of two closest partition functions $Z(J')/Z(J)$ with  $\mathcal{O}(1)$ range (i.e. $0.1\sim10$) is essential--not only to preserve the polynomial computation complexity under importance sampling~\cite{ding2024reweight}, but also to ensure that system error remains controllable (system error analysis can be found in Supplementary Note 7). In addition, the number of slices should be increased near the phase transition to maintain the ratio with $\mathcal{O}(1)$ range, since the distribution of the partition function changes more rapidly in this regime.
Motivated by the reweighting scheme, we propose a novel scheme termed "bipartite reweight-annealing (BRA)"  method to address the challenges of general measurements in QMC simulations. We will present several examples to demonstrate its feasibility and versatility.

\vspace{\baselineskip}
\noindent\textbf{Results}\\
\textit{\color{blue}Bipartite Reweight-annealing Framework.-}
In fact, we realize that the reweighting scheme is not only limited to the standard PF $Z(J)$ but can be applied to any distribution that varies with the related parameters. In practice, an off-diagonal observable can be treated as the ratio of two types of PFs $\langle \mathrm{O} \rangle=\bar{Z}/Z$, where $\bar{Z}=\mathrm{tr} (\mathrm{Oe^{-\beta H}})$. 
This insight inspires us to reweight different kinds of PFs (the numerator $\bar{Z}(J)$ and denominator $Z(J)$) respectively, as Fig. \ref{fig:rean} (b) and (c) show. The key idea is that we firstly calculate the ratios $\bar{Z}(J')/\bar{Z}(J)$ and ${Z(J')}/{Z(J)}$, and if we have a reference point $\bar{Z}(J)/Z(J)$ which is easily solvable (as displayed in Fig. \ref{fig:rean}), then the target measurement $\langle \mathrm{O}(J') \rangle$ can be estimated in this approach:
\begin{equation}
    \langle \mathrm{O(J')} \rangle=\frac{\bar{Z}(J')}{Z(J')}=\frac{\bar{Z}(J)}{Z(J)}\times\frac{\bar{Z}(J')}{\bar{Z}(J)}\times\frac{Z(J)}{Z(J')}
\label{eq:main3}
\end{equation}
where ${\bar{Z}(J)}/{Z(J)}$ is the known reference point, ${\bar{Z}(J')}/{\bar{Z}(J)}$ and ${Z(J)}/{Z(J')}$ can be calculated by reweighting.  Likewise, maintaining the ratio within an $\mathcal{O}(1)$ range is imperative (detail can be found in Supplementary Note 7). 

This BRA scheme avoids the intractable problem of calculating the ratio between two entirely different PFs  (Fig \ref{fig:rean} (a)) by translating it into a solvable framework. It is highly general and can be applied to almost all physical quantities. In the following sections, we will employ this scheme to demonstrate several off-diagonal measurements that previously were rather difficult, even impossible to be calculated in QMC. Moreover, scanning the observables along the path of physical parameter to trace the phase diagram becomes natural and efficient in the BRA frame. Actually, we will show the annealing path is not limited to the physical parameter only, but also works for the degree of freedom in both space and time.

\textit{\color{blue}Equal-time off-diagonal correlations.-}
As an example, we consider the Hamiltonian of the spin-1/2 XXZ model, which is given by:
\begin{equation}
H_{XXZ} =  \sum_{\langle i,j\rangle}\left[ \frac 1 2 (S^+_i  S^-_j + S^-_i S^+_j) + \Delta S^z_i S^z_j \right]  
\label{eq3}
\end{equation}
where $\langle i,j\rangle$ denotes the nearest neighbors, $\Delta$ is the parameter that controls the anisotropy. 
The Hamiltonian can be simulated using the directed loop algorithm of the SSE method (detail can be found in Supplementary Note 1)~\cite{DL1_OlavF_2002,alet2005generalized,DL2_OlavF_2003,syljuaasen2004directed}. In this model,  two-point off-diagonal operators $\langle S_i^x S_j^x \rangle = \langle S_i^y S_j^y \rangle = (\langle S_i^+ S_j^- \rangle + \langle S_i^- S_j^+ \rangle)/4$, which can be measured using worm-like sampling trick 
~\cite{Dorneich2001accessing,wenjing2021measuring,Evertz2003loop}. 
However, measuring a general off-diagonal correlation function is significantly more challenging. 

Here, taking correlation of $S^x$ operators as an example, we show how to measure it via varying the physical parameter $\Delta$ in our scheme,
\begin{equation}
\langle S^x_iS^x_j\rangle_{\Delta} =\frac{\mathrm{tr} (S^x_iS^x_j e^{-\beta H})}{\mathrm{tr} (e^{-\beta H})} = \frac{\bar{Z}(\Delta)}{Z(\Delta)}
\label{eq4}
\end{equation}
where $\bar{Z}(\Delta)$ represents a general partition function with extra off-diagonal operators inserted, distinguished from a normal partition function without these extra off-diagonal operators. The calculation of $\langle S^y_iS^y_j\rangle$ is the same as $\langle S^x_iS^x_j\rangle$ in this frame, as explained in the Supplement note 1.

\begin{figure}[htp]
\centering
\includegraphics[width=0.45\textwidth]{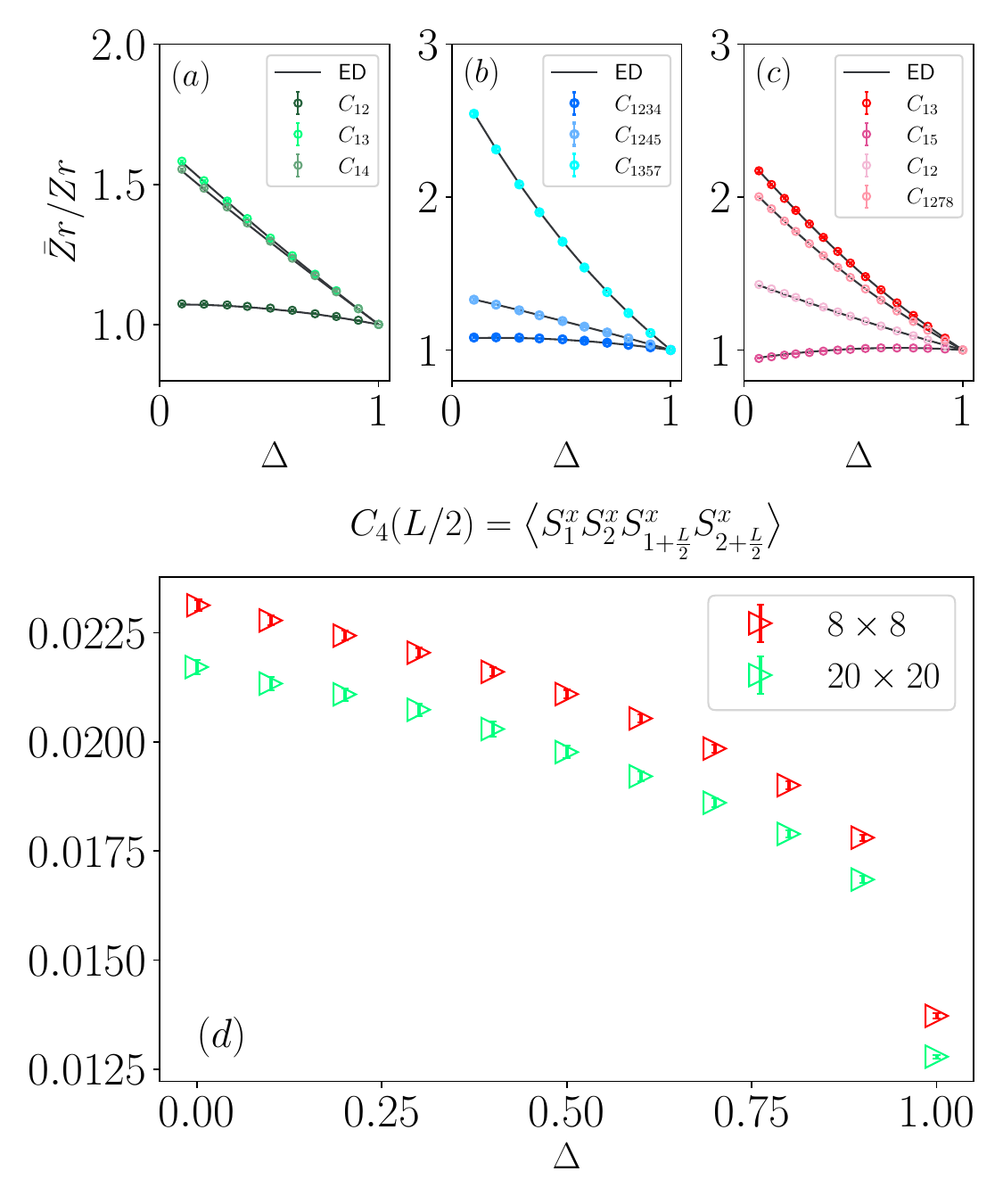}
\caption{Equal-time off-diagonal correlation measurement via the reweight-annealing framework.
(a) Ratios of two-point correlations $C_{ij}=\langle S^x_i S^x_j \rangle$ as a function of the Ising coupling strength $\Delta$ for $L=10$ with $\beta=20$. (b) Ratios of four-point correlations $C_{ijkl}=\langle S^x_i S^x_j S^x_k S^x_l \rangle$ in the one-dimensional chain. (c) Ratios of representative two-point ($C_{ij}$) and four-point ($C_{ijkl}$) correlations on a $4\times 2$ lattice with $\beta=8$. (d) Four-point correlation $C_{12,L/2,L/2+1}=\langle S^x_1 S^x_{2} S^x_{L/2} S^x_{L/2+1}\rangle$ on $8\times 8$ and $20\times 20$ square lattices with $\beta=2L$. Panels (a)-(c) include comparisons with exact diagonalization (ED) results.  Error bars ($\pm 1\sigma$) from SSE simulation denote the standard error of the mean obtained from the Monte Carlo bins.  All calculations are performed on lattices with periodic boundary conditions (PBC).}
\label{fig:xxzbenchmark}
\end{figure}

Firstly, we consider an obvious reference point of this model: $\Delta'=1$, known as the "Heisenberg point" and possessing $\text{SU(2)}$ symmetry.  This implies isotropic spin-spin correlations with $\langle S^z_i S^z_j \rangle = \langle S^x_i S^x_j\rangle$.   Moreover, since $\langle S^z_i S^z_j \rangle$ is accessible through diagonal measurements in a standard QMC framework, the focus shifts to measuring the ratio of the partition functions. For convenience, we define that $\bar{Z}r = \bar{Z}(\Delta)/\bar{Z}(\Delta')$ and $Zr = Z(\Delta)/Z(\Delta')$. 
Then the Eq.(\ref{eq4}) can be rewritten as 
\begin{equation}\label{xxcorr-xxz}
    \langle S^x_i S^x_j\rangle_{\Delta}=\bar{Z}r/Zr\times \langle S^z_i S^z_j \rangle_{\Delta'=1}
\end{equation}
In this way, the correlation of $S^x$ operators can be easily calculated as Fig. \ref{fig:xxzbenchmark} shows (detail can be found in Supplementary Note 2). 

The QMC results are also compared with the exact diagonalization (ED) in order to demonstrate the reliability of this scheme. 
Fig.\ref{fig:xxzbenchmark} shows the calculation results from ED and BRA. The subfigures (a) and (b) exhibit two-point correlations and four-point correlations, represented by $C_2(r) = \langle S_i^x S_{i+r}^x \rangle$ and $C_4 = \langle S_i^x S_j^x S_k^x S_l^x \rangle$ in an XXZ chain with $L = 10$ and $\beta = 20$. Similar simulation results of 2D lattice with $L_x = 4, L_y = 2, \beta = 8$ are shown in the subfigure (c). The black line represents the ED results which match well with the QMC data. We have plotted only a few points on the graph for clarity, while the actual simulation data points of BRA are densely distributed.

One may feel that the $\text{SU(2)}$ symmetry at $\Delta'=1$ is a strict condition which is not general for an arbitrary model. Actually, it is convenient to introduce an auxiliary Hamiltonian $H_0$ with friendly symmetry or easily solvable property. As what quantum annealing does~\cite{das2008colloquium,yan2023quantum,ding2024exploring}, we can set the BRA path as $tH+(1-t)H_0$ and anneal from $t=0$ to $t=1$. This approach allows us to obtain the observable of the target Hamiltonian $H$.

\begin{figure}[htp]
\centering
\includegraphics[width=0.5\textwidth]{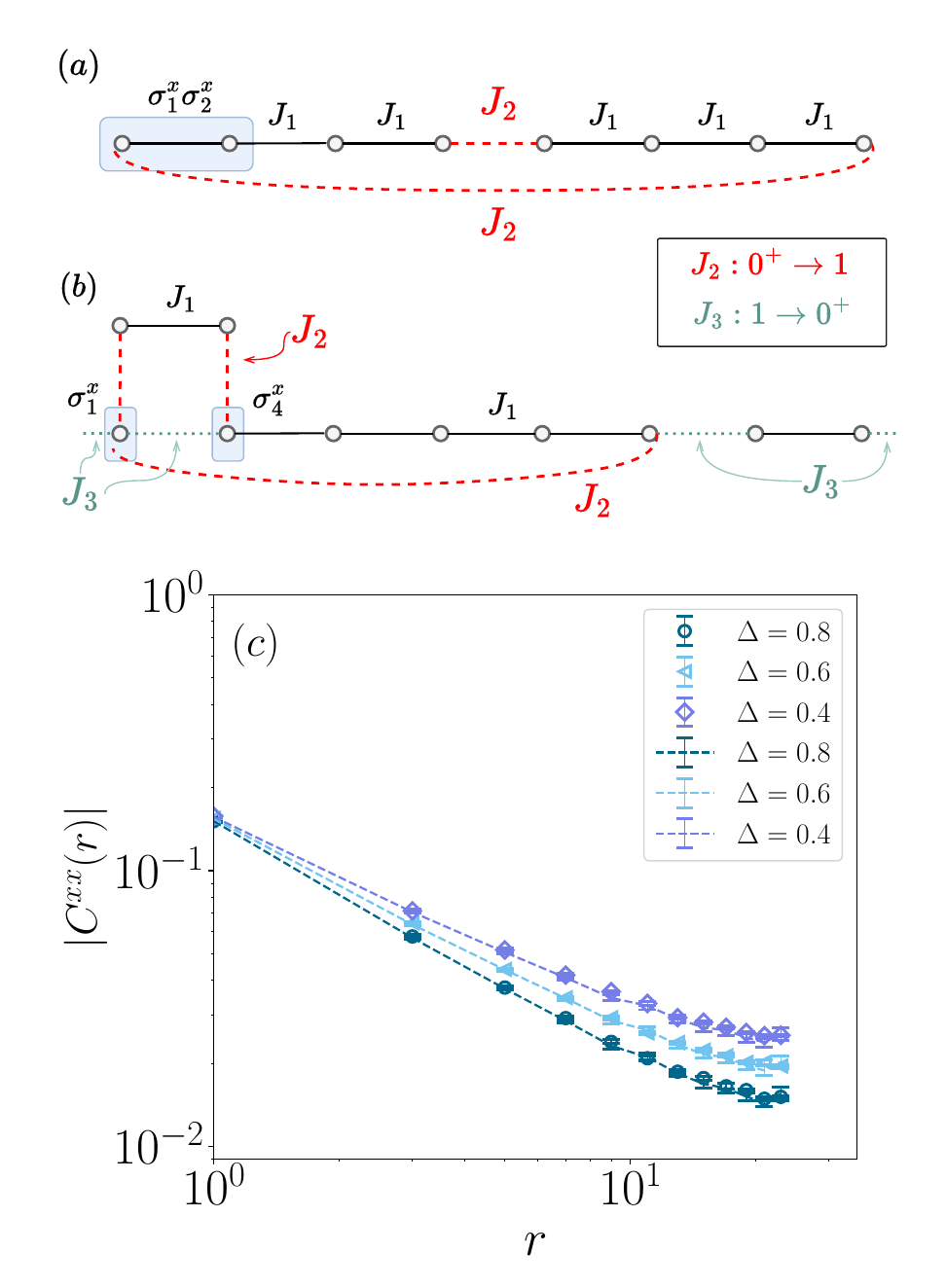}
\caption{ Spatial coupling parameters Annealing.
The off-diagonal correlation measurement via lattice reweight-annealing method in the 1D XXZ model with $L=48$. (a) The lattice diagram for annealing along the system size $L$. We incrementally tune the coupling $J_2$ from $0^+$ to 1. (b) The lattice diagram for annealing along the distance $r$ between $S^x_1$ and $S^x_{1+r}$. We firstly gradually adjust the coupling $J_2$ from $0^+$ to 1, then we gradually tune the coupling $J_3$ from 1 to $0^+$. (c) Two point off-diagonal correlations for system size $L=48$ obtained from subplots $(a)$ and $(b)$ annealing method. The dashed lines represent the simulation for annealing Ising coupling $\Delta$ with fixed $S^x_1S^x_{1+r}$.  Both the hollow symbols and the dashed line represent QMC results, displayed with error bars denoting the mean value and standard error.}
\label{fig:xxzlr1}
\end{figure}

\textit{\color{blue}System Size and Distance Growth by Annealing.-}
Another choice for the reference point is to measure the observable through the ED method in small size, then anneal the small system to large size. 

In this approach, the system size $L$ and distance $r$ between $S^x_1$ and $S^x_{1+r}$ can be considered as BRA parameters. For instance, we can choose the $\langle S^x_1 S^x_2\rangle$ for $L_0=4$ as a reference point, and then we obtain the  $\langle S^x_1 S^x_2\rangle$ for larger system $L$ via adding the remaining sites $L-L_0$ to the original chain, as shown in the Fig.\ref{fig:xxzlr1}(a). In this procedure, the interaction $J_2$ is tuned to couple 4 sites with $L-4$ sites. When we fix the system size and choose the $\langle S^x_1 S^x_2\rangle$ as a reference point, we can obtain the $\langle S^x_1 S^x_{1+r}\rangle$ ($r>1)$ via adding some sites to the area between $S^x_1$ and $S^x_2$ and removing some sites at the end of this chain. We need to tune the coupling $J_2$ from $0^+$ to 1 and also gradually adjust the coupling $J_3$ from 1 to $0^+$, in order to keeping the chain length unchanged as displayed in the Fig.\ref{fig:xxzlr1}(b). A consistent match between the QMC and ED results for off-diagonal correlations is found in Supplementary Note 5.
 This method certainly can be extended to the simulation of large systems. As shown in Fig.\ref{fig:xxzlr1}, we obtain the off-diagonal correlation $C^{xx}(r)=\langle S^x_1 S^x_{1+r}\rangle$ using the reweighting method of changing system size (benchmark with the worm-trick method can be found in Supplementary Note 8). \  In Fig.\ref{fig:xxzlr1} (b), $|C^{xx}(r)|$ has power-law decay, which reflects the physical feature of Luttinger liquid. As $\Delta$ decreases, the power-law parameter becomes smaller, which indicates the $S^x_iS^x_j$ correlation is enhanced and $S^z_iS^z_j$ correlation is weakened. Besides, we utilize the method of annealing $\Delta$, which was introduced in the preceding section, with fixed large-size $L=48$ to obtain the curves of $|C^{xx}(r)|$ (dashed line in Fig. \ref{fig:xxzlr1} (b)), which agrees well with the results through annealing $L$ and $r$.

\begin{figure}[htp]
\centering
\includegraphics[width=0.5\textwidth]{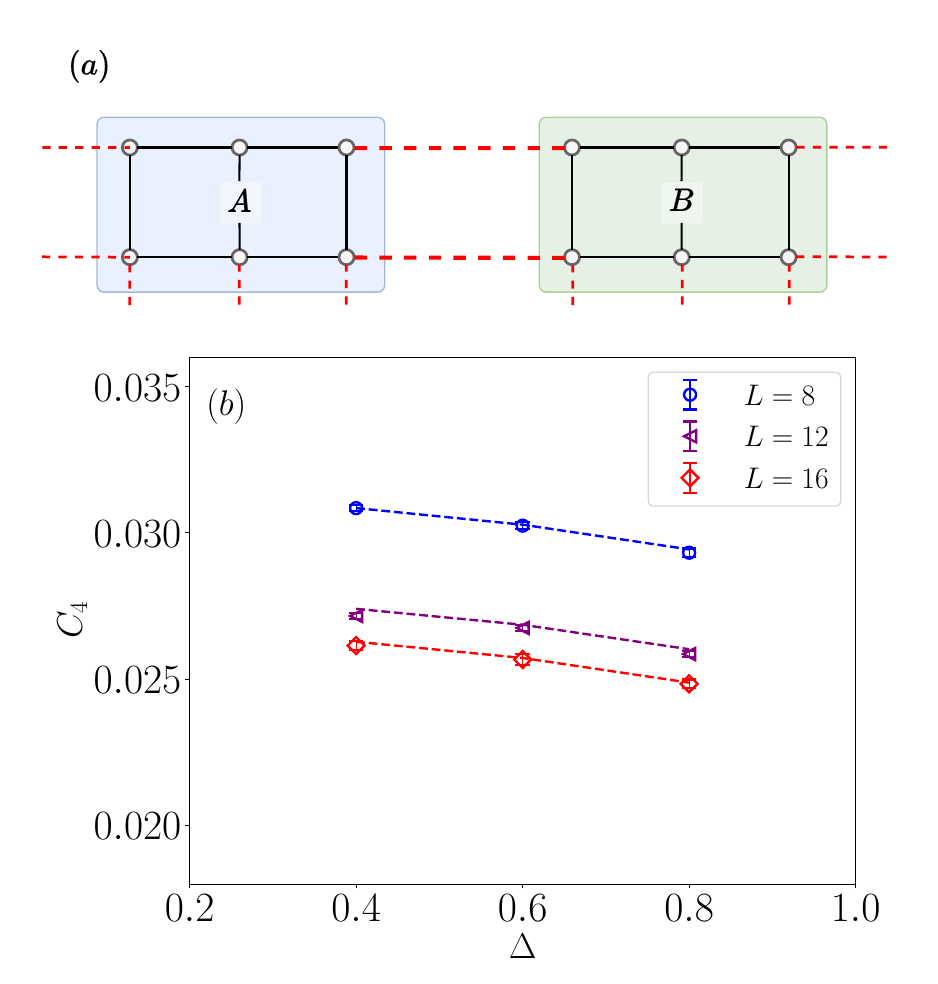}
\vspace{5pt}
\caption{Separability of multipartite off-diagonal observables.
(a) When a large system is decomposed into several parts without coupling, the measured observable can also be separated into the product of independent components. (b) The off-diagonal correlations obtained via the annealing from two small part $A$ and $B$. Here $C_4=\langle S^x_1S^x_2S^x_{L/2+1}S^x_{L/2+2}\rangle$. The $S^x_1S^x_2$ is set on the part $A$, and the $S^x_{L/2+1}S^x_{L/2+2}$ is set on the part $B$ ($i=L/2+1$). The colorful dots are QMC results with standard error as error bars. And the dashed lines are the pure ED results.}
\label{fig:xxzm}
\end{figure}

\textit{\color{blue}Separability of Measurements.-}
The second scheme, which involves annealing from a small system to larger system, inspires us to explore the separability of the general measurement in a large system. 
Without loss of generality, we consider a scenario where a large system is composed of two decoupled smaller subsystems as shown in Fig. \ref{fig:xxzm}. This approach can be easily extended to systems with multiple parts.
In the decoupled case, the density matrix of the total system is the tensor product of the two density matrices, i.e. $\rho=\rho_A\otimes \rho_B$. Typically, we encounter two kinds of measured operators, $O_A\otimes O_B$ and $O_A+O_B$, they satisfy
\begin{align}
        \langle O_A\otimes O_B \rangle_{A\cup B} &=\frac{\mathrm{tr}((\rho_A\otimes \rho_B) (O_A\otimes O_B))}{\mathrm{tr}(\rho_A\otimes \rho_B)} \nonumber\\
        &=\frac{\mathrm{tr}(\rho_A O_A)\mathrm{tr}(\rho_B  O_B)}{\mathrm{tr}(\rho_A) \mathrm{tr}(\rho_B)} \nonumber\\
        &=\langle O_A \rangle_A \langle O_B \rangle_B
\end{align}  
and
\begin{align}
        \langle O_A+ O_B \rangle_{A\cup B} &=\frac{\mathrm{tr}((\rho_A\otimes \rho_B) (O_A+O_B))}{\mathrm{tr}(\rho_A\otimes \rho_B)} \nonumber\\
        &=\frac{\mathrm{tr}((\rho_A\otimes \rho_B)  (O_A \otimes \mathbb{I}_B))}{\mathrm{tr}(\rho_A\otimes \rho_B )} \nonumber\\
        &+\frac{\mathrm{tr}((\rho_A\otimes \rho_B)  (\mathbb{I}_A \otimes O_B)}{\mathrm{tr}(\rho_A\otimes \rho_B )} \nonumber\\
        &=\langle O_A \rangle_A +\langle O_B \rangle_B 
\end{align}
where $\langle ...\rangle_{A\cup B}$ denotes the observable is measured in the total system ${A\cup B}$ and the coupling between $A$ and $B$ is zero. $\langle ...\rangle_{A (B)}$ denotes the measurement in the subsystem $A$ ($B$).

Based on the above two equations, we can firstly decompose a large system into several independent parts without coupling and measure the observables of each part via ED. \  By taking the ED result as a reference point, we then employ QMC to reweight the coupling between each parts from zero to the target value. Consequently, the final observable in the total system can be obtained in this way. 

For example, we assign $S^x_1S^x_2$ operator to subsystem $A$ and another $S^x_iS^x_{i+1}$ operator to subsystem $B$. The expectation value $\langle S^x_1 S^x_2 \rangle_A$ and $\langle S^x_i S^x_{i+1} \rangle_B$ can be obtained via ED since the system size of $A$ or $B$ is small. Subsequently, we incrementally adjust the coupling $J_{AB}$ between $A$ and $B$ to obtain the correlation $\langle S^x_1 S^x_2S^x_i S^x_{i+1} \rangle$. 
As depicted in Fig.\ref{fig:xxzm}, we utilize the above annealing method to obtain the four-point off-diagonal correlation with different system size, and the reference points are obtained with small system size $L'=L/2$ via ED. The QMC results are in excellent agreement with the pure ED results, which demonstrates the reliability of this method.  From a technical standpoint, it is related to the annealing of $L$ and $r$ of the chain in the previous section, but this section more clearly demonstrates the power of this method from small to large sizes. \  In the next section, we will use this approach to calculate disorder operators in 2D systems.

\textit{\color{blue} Disorder operator in 2D TFIM.-}
Here we investigate the off-diagonal measurement for the transverse Ising model (TFIM). The Hamiltonian is given as follows,
\begin{equation}
H_{TFIM} =  -J\sum_{\langle i,j\rangle}\sigma^z_i \sigma^z_j -h\sum_{i}\sigma^x_i
\end{equation}
where $\sigma^{z/x}$ is the Pauli spin-1/2 matrix and $\langle i,j\rangle$ means the nearest-neighbor coupling. $h>0$ is transverse field term and $J>0$ is the ferromagnetic term~\cite{Deng2002Cluster}. 
Because the TFIM only preserves $Z_2$ symmetry, we choose the $J=0^{+}$ and $h=1$ as a reference point. When $J=0$, the reference point $\langle \sigma^x_i\sigma^x_j\rangle=1$ since all the $\sigma^x=1$. In the simulation, we can choose $J\rightarrow 0^{+}$ which makes $\langle \sigma^x_i\sigma^x_j\rangle$ very close to 1. The BRA formula can be expressed as
$\frac{\bar{Z}(J)}{Z(J)}=\bar{Z}r/Zr\times\langle \sigma^x_i\sigma^x_j\rangle_{J=0^{+}}$,
where $\bar{Z}r = \bar{Z}(J)/\bar{Z}(J'=0^+)$ and $Zr = Z(J)/Z(J'=0^+)$. If we want to measure the many-body off-diagonal observables, we just need to change the $\bar{Z}(J)=\langle \sigma^x_i\sigma^x_j\rangle_J$ into $\langle \sigma^x_1\sigma^x_2...\sigma^x_n\rangle_{J}$. 
For TFIM, the QMC results in small system sizes are also well consistent with the ED~\cite{Phillip2017QuSpin,Phillip2019QuSpin}, detail can be found in Supplementary Note 6.

\begin{figure}[htp]
\centering
\includegraphics[width=0.5\textwidth]{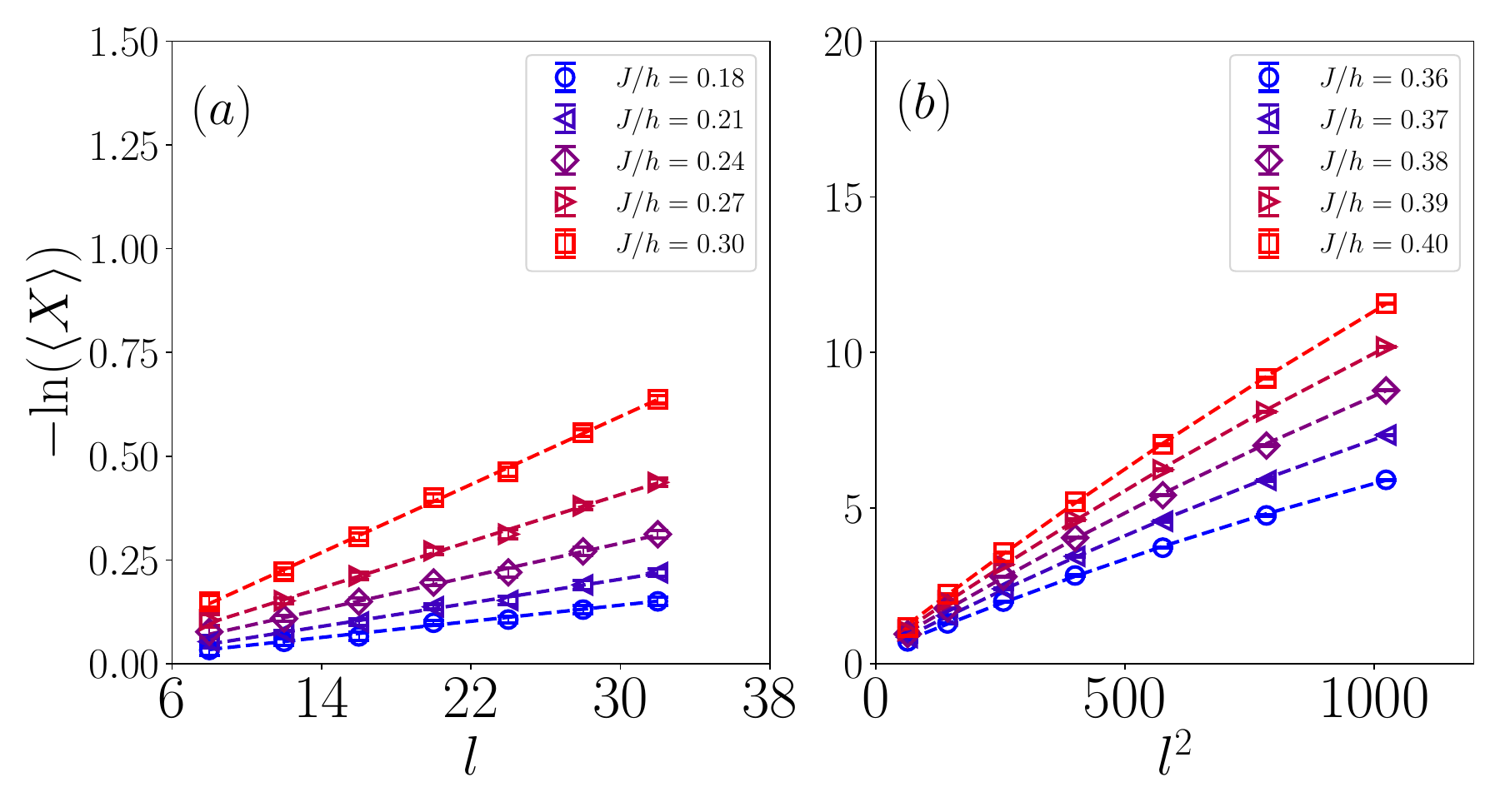}
\caption{The disorder operator $\langle X \rangle$ measurement in the 2D TFIM. 
Setting $L=16$, $\beta=16$ and $h=1$. The dashed lines are the fitting curves. (a) Scaling behaviors of $\langle X \rangle$ in the paramagnetic phase. (b) Scaling behaviors of $\langle X \rangle$ in the ferromagnetic phase. Each error bar represents the statistical one-standard-error estimates.}
\label{fig:disop}
\end{figure}

We then mainly focus on the disorder operator of 2D TFIM on a square lattice. The disorder operator is a non-local operator which can reveal the high-form symmetry breaking and conformal field theory (CFT) information in quantum many-body systems~\cite{zhao2021Higher,Wang2021Scaling,Wang2022Scaling,jiang2023versus,LiuF2023Fermion,liu2023disorder,Liu2024Measuring,Xiao2021Universal,lake2018higherform,fradkin2017disorder,estienne2022cornering}. For 2D TFIM, we define the disorder operator $\langle X\rangle=\langle \prod_{i\in M} \sigma^x_i \rangle$ to detect the non-local information, where $M$ is a $R\times R$ square area in the lattice. Its perimeter is $l=4R$ and it contains $R^2$ off-diagonal operators. This disorder operator, a multi-body off-diagonal observable, was only well measured in the QMC based on $\sigma^x$ basis in the past, which is challenging to obtain directly in the $\sigma^z$ basis~\cite{zhao2021Higher}. Although the operator $\sigma^x$ is contained in the TFIM Hamiltonian and can be measured in the $\sigma^z$ basis in principle~\cite{sandvik1992generalization,Sandvik2003Stochastic}, it suffers from rather large fluctuations due to the requirement of a product of a series of $\sigma^x$ in an area. It requires that the series of $\sigma^x$ operators must appear connectedly in the time-space manifold in the $\sigma^z$ basis, which is a low-probability event.

This difficulty can be overcome via BRA method. As depicted in Fig.\ref{fig:disop}, we have successfully obtained the disorder operator with different perimeters $l$ in the paramagnetic (PM) phase and ferromagnetic (FM) phase (benchmark with the directed calculation in $\sigma_x$ basis can be found in Supplementary Note 9). Here we set $h=1$ and the critical point becomes $J=0.3285$~\cite{Huang2020Worm,zhao2021Higher}. For convenience, we firstly utilize the separability method in the above section to measure the disorder operator at $J/h=0.18$ . Taking it as a reference point, we then obtain the disorder operators for different $J/h$ via annealing along $J$. In the PM phase, the disorder operator satisfies the perimeter law $\langle X \rangle \sim e^{-al}$, which is consistent with the CFT prediction. 
In the FM phase, the disorder operator satisfies the area law $\langle X \rangle \sim e^{-bl^2}$, which reveals the presence of high-form symmetry~\cite{zhao2021Higher}.

\begin{figure}[htp]
\centering
\includegraphics[width=0.5\textwidth]{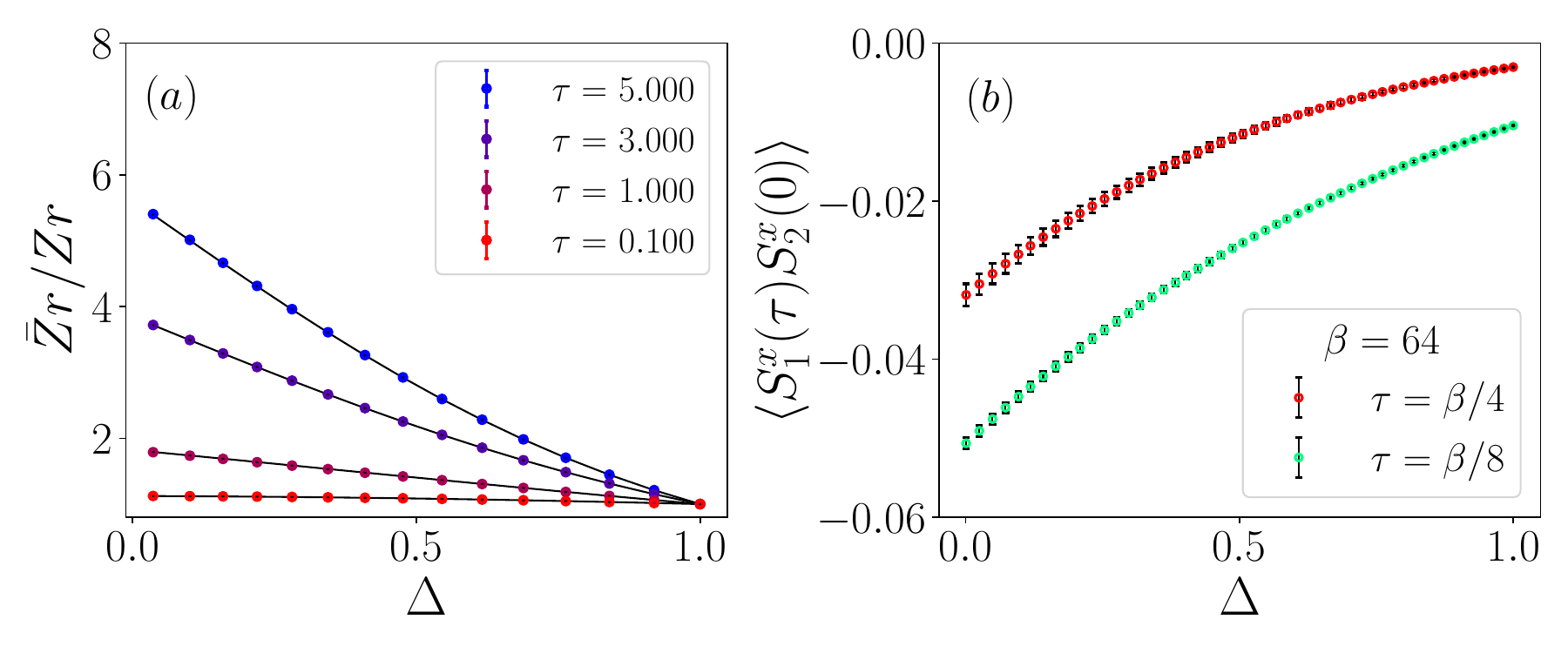}
\caption{The QMC results of the two-point imaginary-time correlation $\langle S^x_i(\tau)S^x_{i+1}(0) \rangle$. $(a)$ The ratio for fixed two-point imaginary-time correlations as the parameter $\Delta$ varies in the XXZ chain, with $L=8$ and $\beta = 10$. For clarity, we plotted only 15 parameter points from the dataset, each matching the ED results (black line). All data points are calculated starting from the Heisenberg condition $\Delta = 1$.     $(b)$ The imaginary-time off-diagonal correlation in the XXZ chain with $L=32$ and $\beta = 64$ for $\tau = \beta/4$ and $\tau = \beta/8$. The associated error bars $(\pm 1\sigma)$ represent the standard error.}
\label{fig:1dxxzImcor}
\end{figure}

\textit{\color{blue}Imaginary-time off-diagonal correlations.-}
Our goal becomes to extend our method to imaginary time correlation functions involving off-diagonal operators. Our discussions will concentrate on the framework of path-integral-like QMC.
The first way based on the physical parameter reweighting is straightforward, which is similar to the method we have employed in the above sections. By fixing two operators at distinct points in imaginary time $\tau$, we have observed the evolution of the imaginary-time correlation function $\langle S_i^x(\tau) S_j^x(0) \rangle$ with varying parameter $\Delta$, as depicted in Fig. \ref{fig:1dxxzImcor}. This is achieved by evaluating the correlation function at several distinct imaginary-time points: $\tau = 0.1$, $\tau = 1.0$, $\tau = 3$, and $\tau = 5$. Notably, when $\beta=10$, $\tau = \beta/2 = 5$ corresponds to the maximum separation in imaginary time. The simulated values, directly comparable as $\bar{Z}r/Zr$, demonstrates excellent agreement with the ED results as shown in the subfigure (a) of Fig.\ref{fig:1dxxzImcor}. For a larger size $L=32$ with $\beta = 64$, as $\Delta$ is tuned from 1 to 0, the imaginary-time off-diagonal correlation $\langle S^x_i(\tau)S^x_{i+1}(0) \rangle$ gradually becomes larger, which is analogous to the equal-time cases.

\begin{figure}[thp]
\centering
\includegraphics[width=0.5\textwidth]{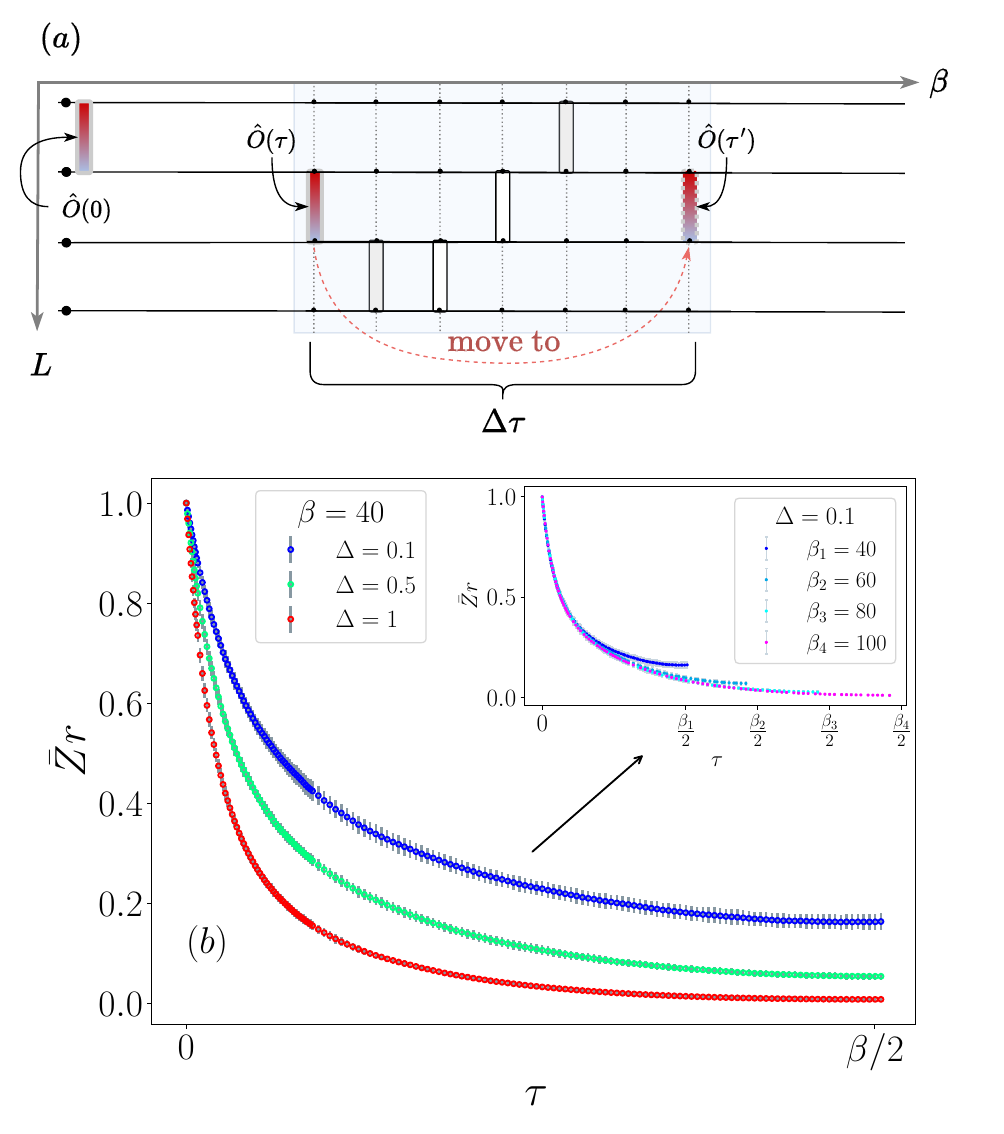}
\caption{The diagram and results of the imaginary-time BRA method. (a) illustrates the schematic of imaginary-time BRA process. The $O$ operators are depicted by a gradient of colors, with one instance inserted and fixed at the imaginary time $\tau=0$, and the other moving within the time axis. If successfully moved, it corresponds physically to a transition from the imaginary time point $\tau$ to a new time point $\tau'$. The time difference is denoted by $\Delta \tau=\tau'-\tau$.    (b) The simulation results for the XXZ chain with $L = 20$ using imaginary-time BRA method are presented. The main plot displays the weight ratios with standard error for a fixed inverse temperature $\beta = 40$ and varying $\Delta$ = 0.1, 0.5, and 1.0. The small inset shows the weight ratios with error bars for fixed $\Delta = 0.1$ and varying $\beta = 2L, 3L, 4L, 5L$. These results have not yet been multiplied by the reference values of $\langle S^x_1(0)S^x_2(0) \rangle$. It can be observed that when $\beta$ is sufficiently large, the furthest correlation $\langle S^x_1(\tau = \beta/2)S^x_2(0) \rangle$ decays to nearly zero.}
\label{fig:imtau_rean}
\end{figure}

Nontrivially, we perform the BRA measurement along the imaginary-time axis, where the distance between two inserted operators increases linearly during the annealing process, as illustrated in Fig.\ref{fig:imtau_rean} (a). For example, we focus on the measurement of $\langle O(\tau)O(0) \rangle$ (the operators are indicated by gradient of colors) currently inserted at time zero and $\tau$, the corresponding partition function for this configuration is $\bar{Z}(\tau)$. We aim to derive $\bar{Z}(\tau')$ for the off-diagonal operator at $\tau'$ using the reweighting technique.
Different from the above schemes for reweighting in which the old/new weight uses a same configuration, the measured operators $O(\tau)$ and $O(\tau')$ represent different configurations here. The solution is to construct an extended ensemble $\bar{Z}(\tau)\cup \bar{Z}(\tau')$, where $\bar{Z}(\tau)$ and $\bar{Z}(\tau')$ are the measured ensembles containing operators $O(\tau)$ and $O(\tau')$, as shown in Fig. \ref{fig:imtau_rean} (a). In this frame, the ratio $\bar{Z}(\tau')/\bar{Z}(\tau)$ can be estimated by the ratio of sampling numbers $N_{\bar{Z}(\tau')}/N_{\bar{Z}(\tau)}$, where the number $N_{\bar{Z}(\tau)}$ or $N_{\bar{Z}(\tau')}$ denotes how many times the sampling belongs to the ensemble $\bar{Z}(\tau)$ or $\bar{Z}(\tau')$. The similar spirit has been used to calculate the entanglement entropy in QMC~\cite{humeniuk2012quantum}. More details about this scheme are explained in the Supplementary Note 3.

\begin{figure}[htp]
\centering
\includegraphics[width=0.5\textwidth]{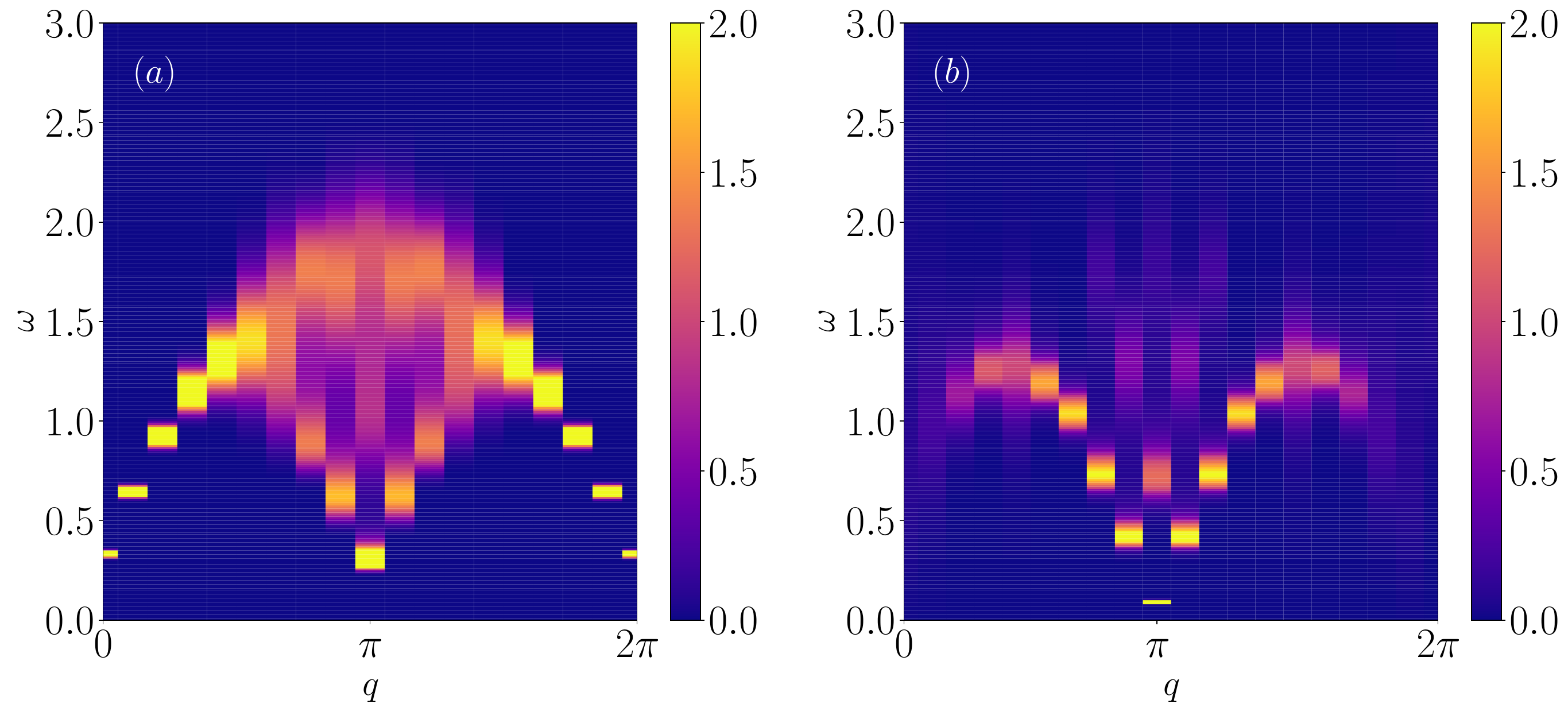}
\caption{The spin excitation spectrum obtained from the SAC method with $L=20$, $\Delta=0.1$ and $\beta=80$. The imaginary-time correlations are extracted from the above BRA method.
(a) The diagonal operator spectrum $S^{zz}(q,\omega)$. (b) The off-diagonal operator spectrum $S^{xx}(q,\omega)$.}
\label{fig:sac}
\end{figure}

We present the numerical outcomes for the XXZ chain in Fig.\ref{fig:imtau_rean} (b). Our analysis has focused on the behavior of the weight ratio $\bar{Z}r$, across three distinct coupling strengths: $\Delta = 0.1$, $\Delta = 0.5$, and the Heisenberg condition $\Delta = 1.0$. We observe that $\bar{Z}r$ initiates from the same starting point for all three curves, with the Heisenberg coupling exhibiting a more rapid decay which reflects the energy gaps in related cases. 
As $\Delta$ decreases, the imaginary-time off-diagonal correlation decays more slowly, indicating the $S^x$ imaginary-time correlation is enhanced that is similar to the equal-time case.
Moreover, the inset illustrates that, at $\Delta = 0.1$, larger $\beta$ makes the ratio $\bar{Z}r$ closer to zero via reducing the finite size effect in imaginary-time direction.

Furthermore, we can obtain the spectrum of operators from the momentum imaginary-time correlations via stochastic analytical continuation (SAC)~\cite{Shao2017Nearly,Sandvik2016Constrained,SHAO2023Progress}. The momentum imaginary-time correlation is defined as $G^{\alpha\alpha}(\mathbf{q},\tau)=\frac{1}{L}\sum_{i,j}e^{-i\mathbf{q}\cdot(\mathbf{r}_{i}-\mathbf{r}_{j})}\langle s^{\alpha}_{i}(\tau)s^{\alpha}_{j}(0)\rangle (\alpha=x,y,z)$. All the real-space off-diagonal imaginary-time correlation can be captured by the above imaginary-time BRA method, which is used to stimulate the excitation spectrum $S^{\alpha\alpha}(\mathbf{q},\omega)$. \  As shown in Fig.~\ref{fig:sac}, the off-diagonal spectrum has sharper lower boundary with weak continuum, which is different from the diagonal spectrum that has strong spinon continuum on the upper boundary (ED results can be found in Supplementary Note 4). Since $\Delta=0.1$ here is close to zero, the difference of the diagonal and off-diagonal spectra
can be understood qualitatively from the limit $\Delta=0$. When $\Delta=0$, the off-diagonal excitation can be solved by the Jordan-Wigner transformation, which is related to a single-mode dispersion of free fermion, thus its excitation is sharp. Meanwhile, the diagonal spectrum $S^{zz}$ corresponds to two fermion operators, which contributes a continuum therefore.
The results demonstrate that our BRA method can be successfully applied to extract the off-diagonal spectrum, which also reveals the different excitation modes compared to the diagonal spectrum with the anisotropic phase. 

\vspace{\baselineskip}
\noindent\textbf{Discussion}\\
In addition to the sign problem in the original ensemble $Z$ (denominator), the numerator $\bar{Z}$ may also exhibit a sign problem. It involves another sign problem within this BRA measurement scheme because we have to calculate the ratio of $\bar{Z}$ with different parameters. For example, when calculating the operator $\sigma^y=-i|\uparrow\rangle\langle \downarrow|+i|\downarrow\rangle\langle \uparrow|$, it introduces an extra sign of $i$ or $-i$ into the weight,  contrasting with the case of $\sigma^x=|\uparrow\rangle\langle \downarrow|+|\downarrow\rangle\langle \uparrow|$. If we attempt to reweight the general PF containing the measured operator $\sigma^y$, denoted as $\bar{Z}_y$, the simulation of ratio $\bar{Z}_y(J')/\bar{Z}_y(J)$ would encounter sign problem. A simple way is to calculate the ratio of $\bar{Z}_y/\bar{Z}_x$, where $\bar{Z}_x$ is the ensemble with the measured operator $\sigma^y$ replaced by $\sigma^x$. Note that $\bar{Z}_y=\sum_i W_i$ and $\bar{Z}_x=\sum_i |W_i|$. As is commonly used in calculating sign value~\cite{ma2024defining,PhysRevB.92.045110,Zhou2019Universal,pan2022sign}, $\bar{Z}_y$ represents the sign system and $\bar{Z}_x$ is the reference system. The ratio $\bar{Z}_y/\bar{Z}_x$ can then be extracted by sampling the reference system $\bar{Z}_x=\sum_i |W_i|$, averaging the sign of each configuration in the sign system ($\bar{Z}_y=\sum_i W_i=\sum_i \mathrm{sign}_i|W_i|$), and ultimately obtaining $\bar{Z}_y/\bar{Z}_x=\langle \mathrm{sign} \rangle$. Finally, the target observable $\bar{Z}_y/Z$ can be derived via $\bar{Z}_y/\bar{Z}_x\times \bar{Z}_x/Z$.

In summary, we propose a variety of detailed schemes in the frame of bipartite reweight-annealing to achieve universal measurement by QMC simulation. Typically, we perform annealing along a physical parameter for the PFs $\bar{Z}(J)$ and $Z(J)$ independently, then connect them via an easily solvable point such as $\bar{Z}(J')/Z(J')$. Thereafter, this concept has been extended to annealing of system size and imaginary time. For example, it is easy to employ ED to calculate the observables in each independent parts and anneal their couplings to construct a large system and solve the target measurement problem. The dynamical behaviors of off-diagonal operators have also been addressed in this work. Off-diagonal spectrum is no longer a natural moat in the quantum many-body computation.
Within this framework, the long-standing problem for the measurement of QMC has been addressed in a general way. 

Essentially, we solve the problem of calculating the overlap between different distribution functions, which is a fundamental challenge in mathematical statistics.
The spirit of BRA can be easily generalized to the measurement of entanglement~\cite{wang2024bipartiterean,ding2024tracking,jiang2024high} and other statistical problems, such as machine learning~\cite{surden2021machine,zhou2021machine,mahesh2020machine}. \\

\vspace{\baselineskip}
\noindent{\bf Methods}
Our work proposes a general measurement framework based on the QMC method, the principle of which has been introduced in the main text. The XXZ model and TFIM model involved employ directed-Loop algorithm and cluster QMC algorithm (detail can be found in Supplementary Note 1 and Note 6), respectively, which will be found in detail in the supplementary information.
\\

\vspace{\baselineskip}

\noindent{\bf Data availability}
The data that support the findings of this study has been archived on Zenodo~\cite{BRA_SSE_code_2025} with the DOI:
\href{https://doi.org/10.5281/zenodo.17669618}{10.5281/zenodo.17669618}.


\vspace{\baselineskip}
\noindent{\bf Code availability}
All code related to this work are available from the authors.

\bibliography{off}

\vspace{\baselineskip}
\noindent{\bf Acknowledgements}
We thank Youjin Deng, Wenan Guo, Yi-Ming Ding and Xuyang Liang for helpful discussions. 
Zenan Liu thanks the China Postdoctoral Science Foundation under Grants No.2024M762935 and NSFC Special Fund for Theoretical Physics under Grants No.12447119. 
Zhe Wang thanks the China Postdoctoral Science Foundation under Grants No.2024M752898. 
This project is supported by the Scientific Research Fund for Distinguished Young Scholars of the Education Department of Anhui Province (No.2022AH020008), the Scientific Research Project (No.WU2024B027) and the Start-up Funding of Westlake University.
The authors thank the high-performance computing center of Westlake University and the Beijing PARATERA Tech Co.,Ltd. for providing HPC resources.

\vspace{\baselineskip}
\noindent{\bf Author Contributions} 
Z.W. (Zhiyan Wang) and Z.L. contribute equally in this work.
Z.Y. initiated the project and conceived the central idea of the BRA algorithm.
Z.W. (Zhiyan Wang) and Z.L. developed the algorithmic implementation, performed the large-scale quantum Monte Carlo simulations, and analyzed the results.
B.B.M carried out the excitation spectrum from exact diagonalization calculations.
Z.W. (Zhe Wang) contributed to the analysis and discussion of the results.
All authors contributed to the interpretation of the findings and to the writing of the manuscript.
Z.Y. supervised the project.


\vspace{\baselineskip}
\noindent{\bf Competing interests}
The authors declare no competing interests.


\clearpage
\appendix
\setcounter{equation}{0}
\setcounter{figure}{0}

\renewcommand{\thefigure}{S\arabic{figure}}
\setcounter{page}{1}

\linespread{1.05}	
\centerline{\bf\Large Supplemental Information} 

\bigskip

Because we are familiar with the QMC-SSE method, the main results are obtained within the framework of SSE simulation. Certainly, this approach is easy to extend to other QMC methods~\cite{prokof1998exact,Nicola2022Efficient,Huang2020Worm,Assaad_book2008}. 

\setcounter{section}{0}
\section*{Supplementary Note 1: Directed-loop update for XXZ model}
\begin{figure}[htp]
\centering
\includegraphics[width=0.5\textwidth]{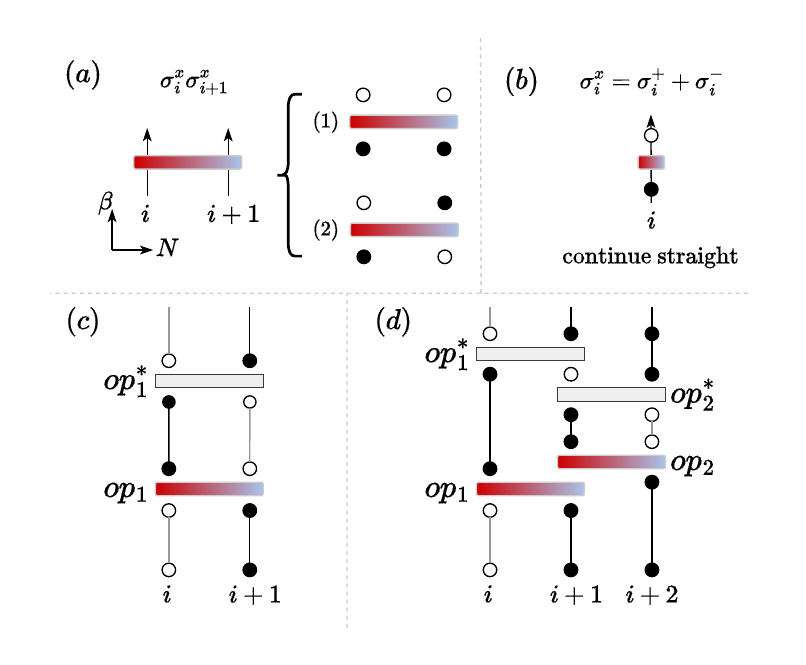}
\caption{The diagram of operators and their update along the path. (a) An off-diagonal $\sigma^x \sigma^x$ operator acts on spins located at lattice sites $i$ and $i+1$. The operator is represented by a gradient-colored bond, where the horizontal axis denotes the direction along the system's lattice sites, and the vertical axis represents imaginary time along $\beta = 1/T$. The arrows indicate the direction of evolution. This operator mainly affects two cases of vertices: (1) vertices where the two spins are parallel before the operator acts, and (2) vertices where the two spins are antiparallel before the operator acts. Here, different spins are represented by black solid circles and hollow circles. Each vertex involves four spins: two spins below the bond (before the operator acts) and two spins above the bond (after the operator acts). (b) A single-site 
$x$-operator action, which directly flips the spin, turning a spin represented by a black solid circle into one represented by a hollow circle. (c) The SSE update involving two two-site operators that affect spins at lattice sites $i$ and $i+1$. Operator $op_1$ is an $\sigma^x \sigma^x$-operator that flips two spins, while $op_1^*$ is an off-diagonal operator in the Hamiltonian (typically $S^+ S^- + S^- S^+$), which also flips two spins. This off-diagonal operator is represented by a gray bond. (d) The updates process with multiple off-diagonal operators, where gradient-colored bonds $op_1$ and $op_2$ represent $\sigma^x \sigma^x$ operators, and gray bonds $op_1^*$ and $op_2^*$ represent off-diagonal operators of type $S^+_i S^-_j + S^-_i S^+_j$. }
\label{fig:QMC}
\end{figure}
In order to realize our idea, we first need to construct the configuration and update of the XXZ model. We employ the previously proposed algorithm for handling the XXZ model, known as the \textit{directed-loop algorithm}\cite{DL1_OlavF_2002,DL2_OlavF_2003,DL3_Henelius2002}. The Hamiltonian of the XXZ model with a longitudinal field, is given by:
\begin{equation}
    H = \sum_{\langle i,j\rangle}\left[ \frac 1 2 (S^+_i S^-_j + S^-_i S^+_j) + \Delta S^z_i S^z_j \right] - h\sum_i S^z_i 
\end{equation}
is divided into the diagonal and off-diagonal operators:  
\begin{align}
H_{1,b} = & \,  [C-\Delta S^z_i S^z_j+h_b(S^z_i+S^z_j)]_b \nonumber\\
H_{2,b} = & \, [\frac{1}{2}(S^+_i S^-_j+S^-_i S^+_j)]_b
\end{align}

In the given system, the parameter $C$ is defined as $C = C_0 + \varepsilon$, where $C_0$ is given by the expression $C_0 = \frac{\Delta}{4} + h_b$. Here, $\Delta$ represents the coupling strength, $h_b$ is from the magnetic field strength defined as $h_b = \frac{h}{2d}$, with $h$ being the magnetic field and $d$ is the dimension of the system. The term $\varepsilon$ is a constant added to ensure that the weights remain positive. The symbols here are consistent with those in the Ref. \cite{DL1_OlavF_2002} to avoid confusion. For simplicity in our discussion, we set $h_b = 0$ and leave $\varepsilon$ unchanged, for example, $\varepsilon = 1$ (these settings are not strictly necessary).  The index $b$ signifies the location of a two-point interaction bond within the system. In this framework, operators are considered as vertices, with each vertex possessing four "legs" that represent the state of two spins before and after the operator's action. This concept involves six types of vertices originally present in the directed loop algorithm, which are expressed as:
\begin{align}
\langle \uparrow \uparrow  |H_{1,b}|\uparrow \uparrow  \rangle  &= \varepsilon,     \langle \downarrow \downarrow   |H_{1,b}|\downarrow \downarrow   \rangle = \varepsilon,\quad (h_b = 0) \nonumber \\
  \langle \uparrow \downarrow |H_{1,b}|\uparrow \downarrow \rangle  & =  \langle  \downarrow\uparrow |H_{1,b}| \downarrow \uparrow\rangle =\Delta/2+\varepsilon,   \nonumber \\
  \langle \downarrow \uparrow  |H_{2,b}|\uparrow \downarrow \rangle & =    \langle \uparrow \downarrow   |H_{2,b}|\downarrow \uparrow  \rangle = \frac{1}{2}
\label{eqDLvertex}
\end{align}
where the $\ket{\uparrow}$ and $\ket{\downarrow}$ represent the states of the spin in the $S^z$-basis. 

\begin{figure}[htp]
\centering
\includegraphics[width=0.5\textwidth]{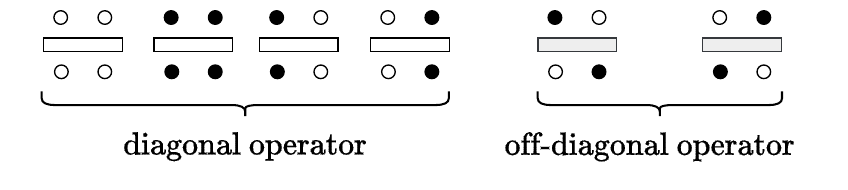}
\caption{Six vertices of the directed loop algorithm designed for the XXZ model without transverse field. The black solid circle and holes represent two types of spins respectively. The first four vertices correspond to diagonal operators, which only contribute to the weight provided by Eq.(\ref{eqDLvertex}) on the two spins and do not flip the spins. The latter two vertices correspond to off-diagonal operators, which flip the spin and contribute to the weight given in Eq.(\ref{eqDLvertex}). Here, blank bonds and gray-filled bonds are used to distinguish the diagonal operators from the off-diagonal operators. }
\label{fig:vertex}
\end{figure}

As Fig. \ref{fig:vertex} shown, different spins are represented by empty circles and black solid circles, with Eq.(\ref{eqDLvertex}) detailing the weights associated with these vertices. Building upon the original six vertex types of the directed loop algorithm, we introduce an additional vertex type, denoted as $\sigma^x \sigma^x$, which is the measured operator.  (It is need to note that even though the Hamiltonian represents a $S = 1/2$ system, we can still choose to insert the Pauli operator for practice reason rather than the $S = 1/2$ operator. This is because the corresponding factor (if it is a two-point correlation operator, then $S^x S^x = 1/4 \times \sigma^x \sigma^x$) will cancel out in both the numerator and denominator of Eq.(\ref{eqs4}):
\begin{equation}
\langle S^x_iS^x_j\rangle_{\Delta} =\frac{\mathrm{tr} (S^x_iS^x_j e^{-\beta H})}{\mathrm{tr} (e^{-\beta H})} = \frac{\bar{Z}(\Delta)}{Z(\Delta)}
\label{eqs4}
\end{equation}
The measurement quantity depends on the reference point and thus does not conflict with the symbol $S^x$ in the text.

The new vertex type can be thought of as consisting of two individual $\sigma^x$ operators that act to flip the spins. It is capable of interacting with any spin states, including those where the two spins are parallel.  As depicted in Fig. \ref{fig:QMC} (a), this operator's sole function during the update is to simultaneously flip any state of the two spins, as demonstrated in cases (1) and (2), where both spins are flipped in two distinct patterns. Case (1) illustrates its action on two parallel spins, while case (2) shows its action on two antiparallel spins. Fig.\ref{fig:QMC} (b) further illustrates the process of spin flipping.

For the off-diagonal operator $S^+_iS^-_j + S^-_iS^+_j$ extracted from the Hamiltonian, it has a weight of zero when applied to two parallel spins. There are only three valid update pathways: "bounce" , "switch-and-continue", and "switch-and-reverse". The latter two pathways permit the transformation of this operator into a two-body diagonal operator. However, the $\sigma^x \sigma^x$ off-diagonal operator, which is central to our measurement, allows for the directed loop's arrow to pass through one of the spins, enabling a continue-straight update. By limiting the operator to this singular update pathway, we prevent it from evolving into a diagonal form or shifting its position during the update process.

To establish a manifold corresponding to $\bar{Z}$ with inserting such an operator, we begin by considering the periodic boundary conditions in the imaginary-time direction. This configuration enables us to insert the operator at any desired layer along the $\beta$-axis in imaginary time. As shown in Fig. \ref{fig:QMC} (c), we denote the inserted $\sigma^x \sigma^x$ operator with the symbol $op$.   It is essential to know that we need to simultaneously introduce another off-diagonal operator (denoted as $op^*$ for distinction) to ensure that the spin on the same site undergoes an even number of flips, in order to satisfy the conservation law \cite{DL2_OlavF_2003}. And it is not necessary to worry about the additional $op^*$ introduced, as its position in imaginary time will adjust with each update.  

In summary, for the measured operators, the update-lines just go straight to cross them which keeps them unchanged. For the normal operators, we keep the original update scheme as usual.

\begin{figure}[htp]
\centering
\includegraphics[width=0.5\textwidth]{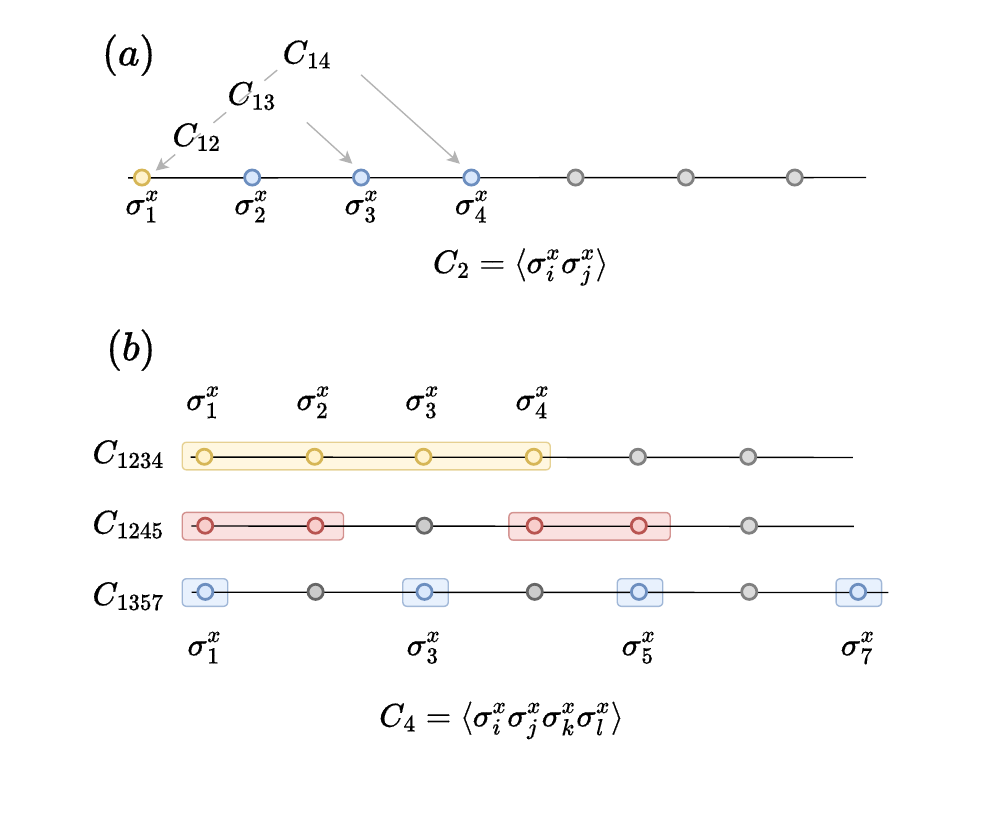}
\caption{Illustration of the measurement schemes for two-point and four-point correlations in the XXZ chain with $L=10$. In all cases, light-colored sites indicate the positions where the $S^x$ operators are applied. (a) Two-point correlation $C_2=\langle S^x_i S^x_j \rangle$. (b) Four-point correlation $C_4=\langle S^x_i S^x_j S^x_k S^x_l \rangle$, with several representative operator placements shown.}
\label{fig:1dxxzchain}
\end{figure}

\begin{figure}[htp]
\centering
\includegraphics[width=0.5\textwidth]{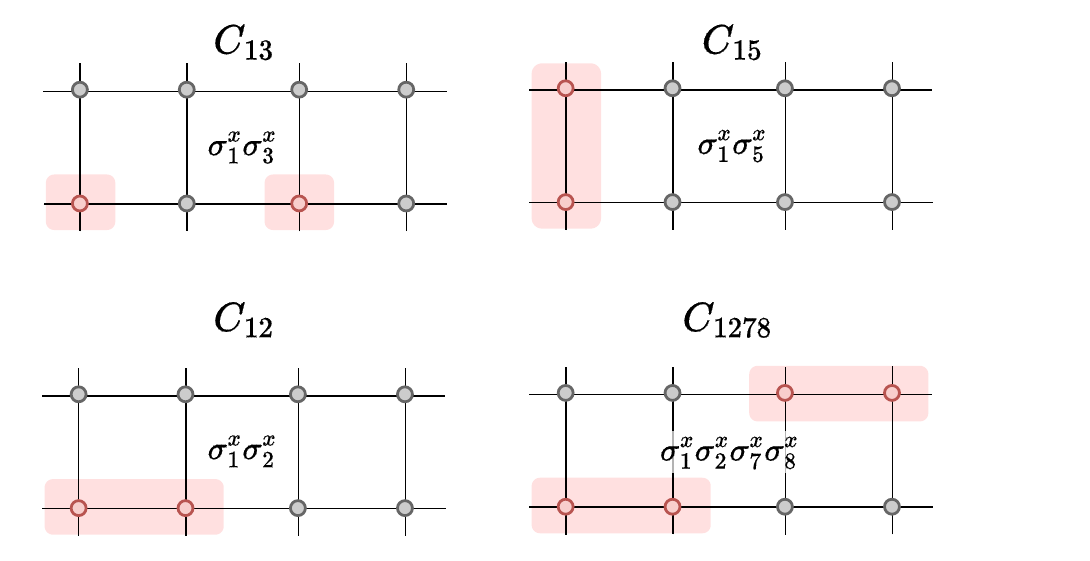}
\caption{Measurements of two-point ($C_2$) and four-point ($C_4$) correlations in the $4\times 2$ XXZ model with $\beta=8$. The operator placements for evaluating these correlations are indicated by the color blocks in the lattice.}
\label{fig:2dxxzlattice}
\end{figure}

\section*{Supplementary Note 2: Observables} 

The insertion of $\sigma^x \sigma^x$ operator is not limited to spin-1/2 system, but also can be applied to other spin systems.
Moreover, it does not impact the observables.   Previous research~\cite{ding2024reweight} has established that the ratio of the partition functions for two points with a varying parameter in SSE. Here, we illustrate its application in the XXZ model.

The partition function is derived by summing the weights of all permissible configurations within a $(d+1)$-dimensional space, $Z = \sum_C W(C)$, where $W(C)$ denotes the weight of a specific configuration $C$. For a bipartite or unfrustrated system, it can be expanded as:
\begin{align} 
Z&=\sum_{\alpha} \sum_{S_{M}} W\left(\alpha, S_{M}\right) \nonumber \\
W\left(\alpha, S_{M}\right) &=    \frac{\beta^{n}(M-n)!}{M!}\langle\alpha| \prod_{p=0}^{M-1} H_{a(p), b(p)}|\alpha\rangle
\label{eq:ra1}
\end{align}
where $n$ represents the number of non-identity elements, or non-unit operators, within the fixed-length operator string $S_M$, and is less than the series cut-off $M$. Here, $\beta$ signifies the inverse temperature,  $\alpha$ refers to the inserted complete basis, and the Hamiltonian is a sum of local operators, such as the six listed in Fig.\ref{fig:vertex} and Eq.(\ref{eqDLvertex}). These operators randomly appear between states, with each occurrence counted as $n_{op}$, where $n_{op} = \{ n_{\downarrow \downarrow}, n_{\uparrow \uparrow}, \ldots \}$. Thus, the total weight of a configuration $C$ corresponds to the individual operator weights and their powers, multiplied by a factor.

To obtain the partition function for a configuration under parameter $J_2$ from that under $J_1$, the ratio of the partition functions is given by:
\begin{align}
\frac{Z\left(J_2\right)}{Z\left(J_1\right)} &=\frac{1}{Z\left(J_1\right)}\bigg ( \sum_C W(C;J_2) \bigg) \nonumber \\
  & =\frac{1}{Z\left(J_1\right)}\bigg ( \sum_C W(C;J_1)  \frac{ W(C;J_2)}{ W(C;J_1)} \bigg)
\label{eq:ra2}
\end{align}

Here, $\frac{W(C; J_2)}{W(C; J_1)}$ is considered as the operator $\hat{R}$ to be measured under $J_1$, such that,
\begin{equation}
    \langle \hat{R} \rangle_{J_1} = \frac{\sum_C \hat{R}(C) W(C;J_1)}{Z\left(J_1\right)} = \left\langle \frac{W(C; J_2)}{W(C; J_1)} \right\rangle_{J_1}
\label{eq:ra3}
\end{equation}
which is in accordance with $\frac{Z(J')}{Z(J)}=\bigg\langle \frac{W(J')}{W(J)} \bigg\rangle_{J}$. Furthermore, for a single permissible configuration under $J_1$, if $J_2 = \Delta_2$ and $J_1 = \Delta_1$, the configuration's weight can be expanded as:

\begin{equation}
    \frac{W(C; \Delta_2)}{W(C; \Delta_1)} = \left( \frac{\Delta_2/2 + \varepsilon}{\Delta_1/2 + \varepsilon} \right)^{n^{\text{diag}}_{\uparrow \downarrow; \downarrow \uparrow}} \times \left( \frac{\varepsilon}{\varepsilon} \right)^{n_{\downarrow \downarrow; \uparrow \uparrow}} \times \left( \frac{1/2}{1/2} \right)^{n^{\text{off-diag}}_{\uparrow \downarrow; \downarrow \uparrow}}
\label{eq:ra4}
\end{equation}

In this expression, the factor in the weight $W(C)$ has been canceled out, and we have utilized the operator weights provided by Eq.(\ref{eqDLvertex}). For Eq.(\ref{eq:ra4}), the remaining terms are:
\begin{equation}
    \left\langle \frac{W(C; \Delta_2)}{W(C; \Delta_1)} \right\rangle_{J_1} = \left\langle \left( \frac{\Delta_2/2 + \varepsilon}{\Delta_1/2 + \varepsilon} \right)^{n^{\text{diag}}_{\uparrow \downarrow; \downarrow \uparrow}} \right\rangle_{\Delta_1}
\label{eq:ra5}
\end{equation}

This demonstrates that the measurement essentially involves counting the number of times the diagonal operator acts on spins that are antiparallel to each other. Another manifold consistently inserts an operator $O$, and the derivation of its partition function $\bar{Z}$ is analogous to Eq.(\ref{eq:ra2})-(\ref{eq:ra5}). Ultimately, to estimate the expectation value of the operator $O$ at the parameter point $\Delta_2$, the ratio of the two partition functions is linked, that is,

\begin{align}
\langle O \rangle_{\Delta_2} = & \frac{\bar{Z}(\Delta_2)}{Z(\Delta_2)}  \nonumber\\
\nonumber\\
= & \frac{\bar{Z}(\Delta_2)}{\bar{Z}(\Delta_1)} \bar{Z}(\Delta_1)  \times \frac{Z(\Delta_1)}{Z(\Delta_2)}\frac{1}{Z(\Delta_1)}\nonumber\\
 = & \frac{\bar{Z}(\Delta_2)}{\bar{Z}(\Delta_1)}\times \frac{Z(\Delta_1)}{Z(\Delta_2)} \times \underbrace{\frac{\bar{Z}(\Delta_1)}{Z(\Delta_1)}}_{\text{reference point}} \nonumber\\
 = & \bar{Z}_r / Z_r \times \langle O\rangle_{\Delta_1} \nonumber\\ 
\label{eq:ra5}
\end{align}

It can be readily extended to multiple parameter points $\Delta_3, \Delta_4, ... \Delta_k$, as depicted in the main text, where,
\begin{equation}
    \langle \mathrm{O(J')} \rangle=\frac{\bar{Z}(J')}{Z(J')}=\frac{\bar{Z}(J)}{Z(J)}\times\frac{\bar{Z}(J')}{\bar{Z}(J)}\times\frac{Z(J)}{Z(J')}
\label{eq:main3}
\end{equation}
and here the paratemeter $\Delta$ plays the role of $J$. The reference point is typically chosen to be the expectation values of operators that are easy to solve. And we denote the product of the specific ratios of partition functions during the annealing process as follows:
\begin{align}
    \bar{Z}_r & = \prod_{i>1}\bigg \langle \left( \frac{ (\Delta_i/2+\varepsilon)}{ (\Delta_1/2+\varepsilon)} \right)^{n_{\uparrow \downarrow;\downarrow \uparrow}} \bigg \rangle_{\bar{Z}}, \nonumber\\
    Z_r  & = \prod_{i>1}\bigg \langle \left( \frac{ (\Delta_i/2+\varepsilon)}{ (\Delta_1/2+\varepsilon)} \right)^{n_{\uparrow \downarrow;\downarrow \uparrow}} \bigg \rangle_Z
\label{eq:ra6}
\end{align}

In order to compare with the ED results, we initially perform the calculation in the small size, as shown in Fig. \ref{fig:1dxxzchain} and \ref{fig:2dxxzlattice}, which examines the one-dimensional and two-dimensional XXZ model respectively. Let's consider the one-dimensional system as an example; the process for two dimensions is analogous. For the one-dimensional case, measuring the nearest-neighbor correlation is straightforward; it simply involves directly inserting a two-body operator at the lattice site pair $(i,i+1)$, see Fig. \ref{fig:QMC} (c). To assess the next-nearest-neighbor correlation, one can either place a single $\sigma^x$ operator at the positions $i$ and $i+2$, or insert the two-body operator at the positions $(i,i+1)$ and $(i+1,i+2)$ respectively, as illustrated in Fig. \ref{fig:QMC} (d).  Because the spin of site $i+1$ has been flipped twice, it satisfies the requirement of imaginary-time periodic boundary conditions. Therefore, as long as there are no other operators between $op_1$ and $op_2$, the $\sigma^x$-operator can effectively be regarded as not acting on the $i+1$ site during the BRA process.  The benchmark results, including both non-nearest neighbor correlators and multi-point correlators, are presented in the main text.

\begin{figure}[htp]
\centering
\includegraphics[width=0.5\textwidth]{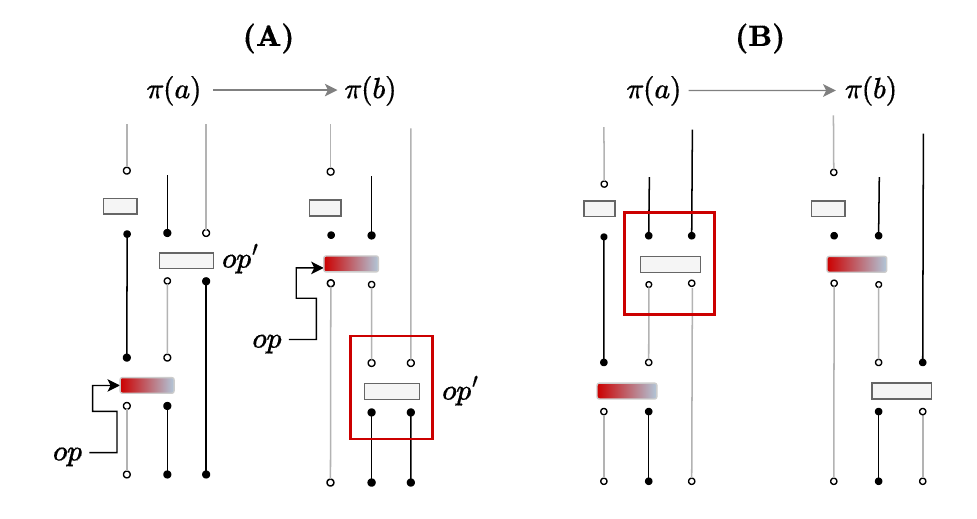}
\caption{Two cases of invalid $op$-moves are presented.  (A). The configuration $a \to b$. Above the operator $op$, there is an off-diagonal operator $op'$ involving one-sided interactions, which can only act on two anti-parallel spins. Suppose that $op$ is moved to a higher level than $op'$. Since the previous spin states are already determined, $op'$ has to act on two parallel spins now. In this case, the vertex associated with $op'$ is invalid, and its contribution of weight is zero.   (B). When the current configuration is still $\pi(a)$, the weight of the configuration is zero. However, after the exchange, the resulting configuration becomes a valid configuration, thereby contributing a non-zero weight.}
\label{fig:vality}
\end{figure}

\section*{Supplementary Note 3: Imaginary-time correlation BRA} 
In the SSE framework, considering the continuous limit, which implies the cut-off $M \to \infty$, the series index $p$ along the $\beta$ evolution direction has a simple correspondence with the path integral imaginary time slices, that is

\begin{equation}
    \tau = \frac{p \beta}{M}
\end{equation}
which only requires that $M$ be sufficiently large. Even for finite $M$, the series index $p$ still correlates with the distribution of imaginary time~\cite{sandvik2019stochastic,Sandvik2010Computational}. Consequently, operator insertion layers, including those without insertions which are treated as identity operator layers, are mapped to their respective imaginary time points in practical computational transformations. 

In the main text, we presented measurements of imaginary-time correlations, which are not significantly different from the equal-time correlations introduced before. Considering an operator inserted at imaginary time $\tau_1$ and another at $\tau_2$, the only requirement is to keep these two operators unchanged while adjusting the system parameters. When these two operators are positioned close enough in imaginary time such that no other operators can insert between them, the situation effectively reduces to equal-time correlations. An interesting case arises when we fix the parameters and perform BRA along the $\beta$-direction. 

\begin{figure}[htp]
\centering
\includegraphics[width=0.5\textwidth]{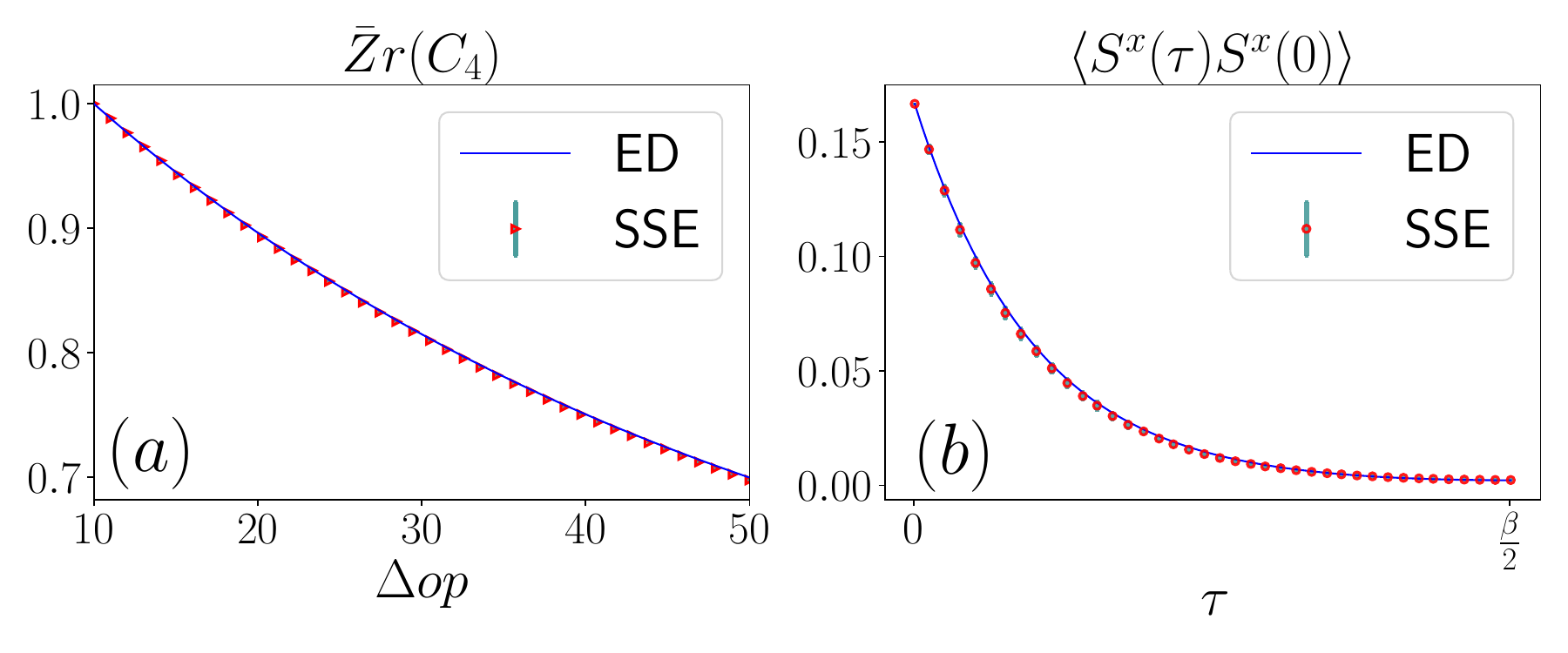}
\caption{BRA along with the $\beta$-direction result of  XXZ chain with $L = 4$ and $\beta = 10$. $(a)$ The weight ratio of the 4-point imaginary-time correlation function $\langle S^x_1(\tau) S^x_2(\tau) S^x_3(0) S^x_4(0) \rangle$.   $\bar{Z}r$ represents the ratio of occurrences between adjacent layers during the BRA process of the inserted operator which corresponds to the weight ratio.  As one two-point $S^x S^x$ operator is fixed at the 0th layer in the $\beta$-direction, the second two-point operator is moved from the 10th layer to the 50th layer.  $(b)$ The two-point imaginary-time correlation $\langle S^x_1(\tau) S^ x_2(0)  \rangle$. The value of $\tau = \beta$ is divided into 1000 layers, with 500 layers annealed from 0.  Error bars denote the standard error on the mean.}
\label{fig:imtaure-ansmall}
\end{figure}

For the transition from $\tau_{i}$ to the next $\tau_{i+1}$, it can be visualized as the movement of one of the $\sigma^x \sigma^x$, denoted as $op$. Taking the movement of a single layer as an example, each movement can be considered as an exchange with the operator $op'$ above $op$. This approach is straightforward if $op'$ is either a diagonal operator or identity operator. However, a significant challenge emerges in certain configurations. After the exchange, configurations that initially had zero weight might now have non-zero weight. This suggests that by remaining in the current configuration, we may be missing out on sampling certain configurations, thereby rendering the sampling process inefficient.  \    As shown in Fig. \ref{fig:vality} (A), the current configuration is in $\pi(a)$ with the inserted operator at $\tau_i$. When we attempt to move the inserted operator to $\tau_i + 1$, the configuration shifts to $\pi(b)$. However, due to the presence of a normal off-diagonal operator at the $\tau_i + 1$ layer, this movement renders the off-diagonal operator invalid after the exchange. Note that Fig. \ref{fig:vality} (B) is not illustrating a reverse process but rather the transition from $\pi(a)$ to $\pi(b)$. The configuration $\pi(a)$ has zero weight due to invalid off-diagonal operator acting on parallel spins, yet it can still result in a valid, non-zero-weight configuration through exchange. Therefore, the detailed balance between $\pi(a)$ and $\pi(b)$ must still be taken into account.

We carry out the exchange with the true probability and then count the number of exchanges and non-exchanges. More precisely, we count how many times the inserted operator appears at $\tau_{i}$  and $\tau_{i+1}$, which provides us the ratio of weight:
\begin{equation}
    \bar{Z}r= \frac{\bar{Z}(\tau _{i +1})}{\bar{Z}(\tau_i)} =  \bigg \langle \frac{N(\tau_{i+1})}{N(\tau_i)} \bigg \rangle
\end{equation}

This counts within the same manifold. The QMC results for smaller system sizes are presented in Fig. \ref{fig:imtaure-ansmall}, showing excellent agreement with the ED results.   More efficient choices are necessary for larger sizes. For the main text, during the simulation, when the imaginary time intervals are relatively small, the operator moves by  \ 10 or \ 20 layers per \  Monte Carlo (MC) step, and for larger intervals, it moves by 50 or even 100 layers, enabling us to efficiently capture the behavior of the imaginary-time off-diagonal correlations as a function of $\tau$.

\section*{Supplementary Note 4: Spectral Function of XXZ model}

\begin{figure}[htp]
\centering
\subfigure[]{

 \includegraphics[width=0.4\linewidth]{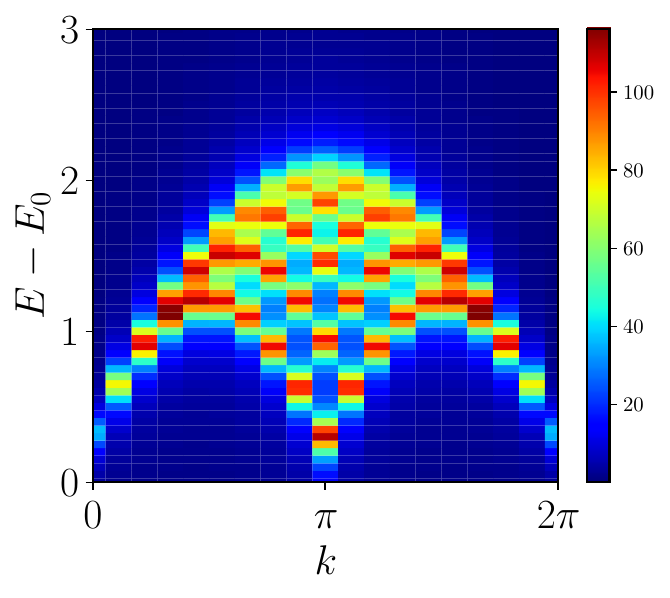}
}
\subfigure[]{
 \includegraphics[width=0.42\linewidth]{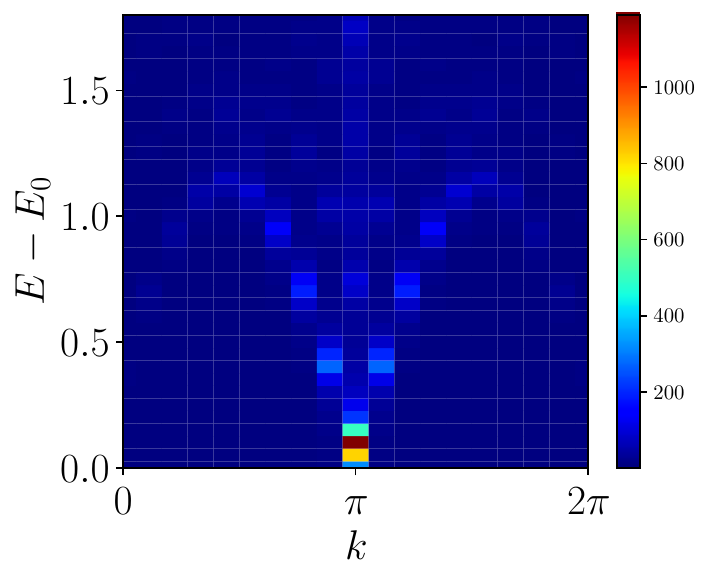}
}
\caption{ED results of spectral function $S^{zz}(q,\omega)$ (a)  and $S^{xx}(q,\omega)$ (b) for 1D XXZ chain with $\Delta=0.1$.}
\label{Specfunc}
\end{figure}
In order to benchmark the excitation spectrum from QMC simulation, we directly use ED to calculate the spectrum function. The spectrum function of target operator $O$ can be defined as,

\begin{align}
S(\omega)=\frac{1}{\pi}\sum\limits_{m,n}e^{-\beta E_n}|\langle m|O|n\rangle|^2\delta(\omega-[E_m-E_n])
\label{eq15-1}
\end{align}

In the ED calculation, we choose the operator $O$ as $O=S^{\alpha}_{q}$ ($\alpha=x,y,z$), where $S^{\alpha}_{q}=\frac{1}{\sqrt{N}}\sum\limits_{i} e^{-ikr_i}S^{\alpha}_{i}$. The exictation spectra from ED are consistent with the results from QMC (see Fig.~6 in the main text).

\section*{Supplementary Note 5: Annealing $L$ and $r$ for XXZ model}

\begin{figure}[htp]
\centering
\includegraphics[width=0.5\textwidth]{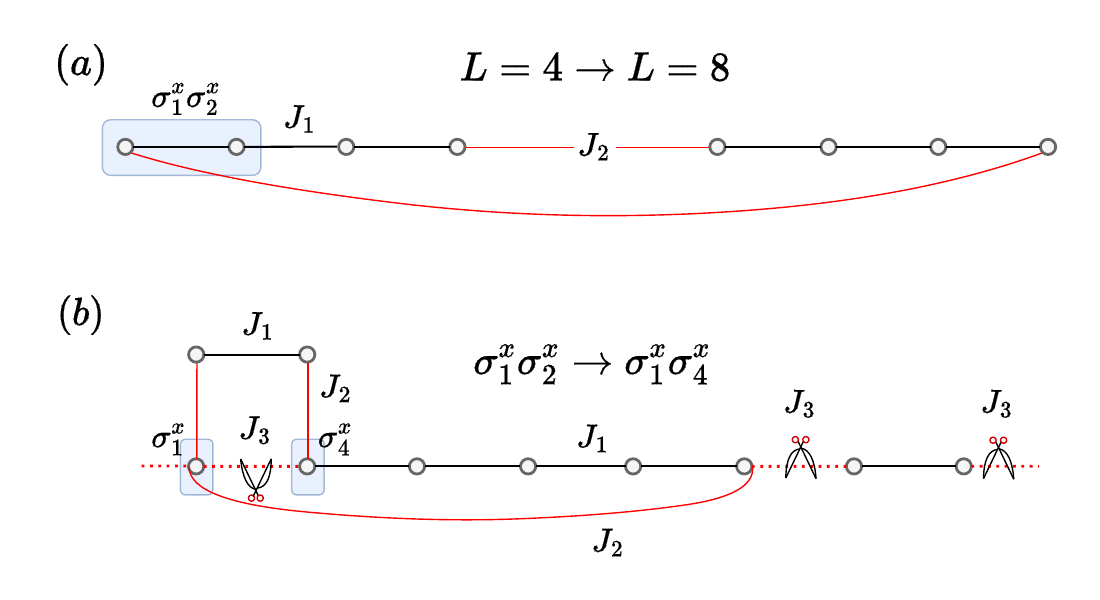}
\vspace{5pt}
\includegraphics[width=0.46\textwidth]{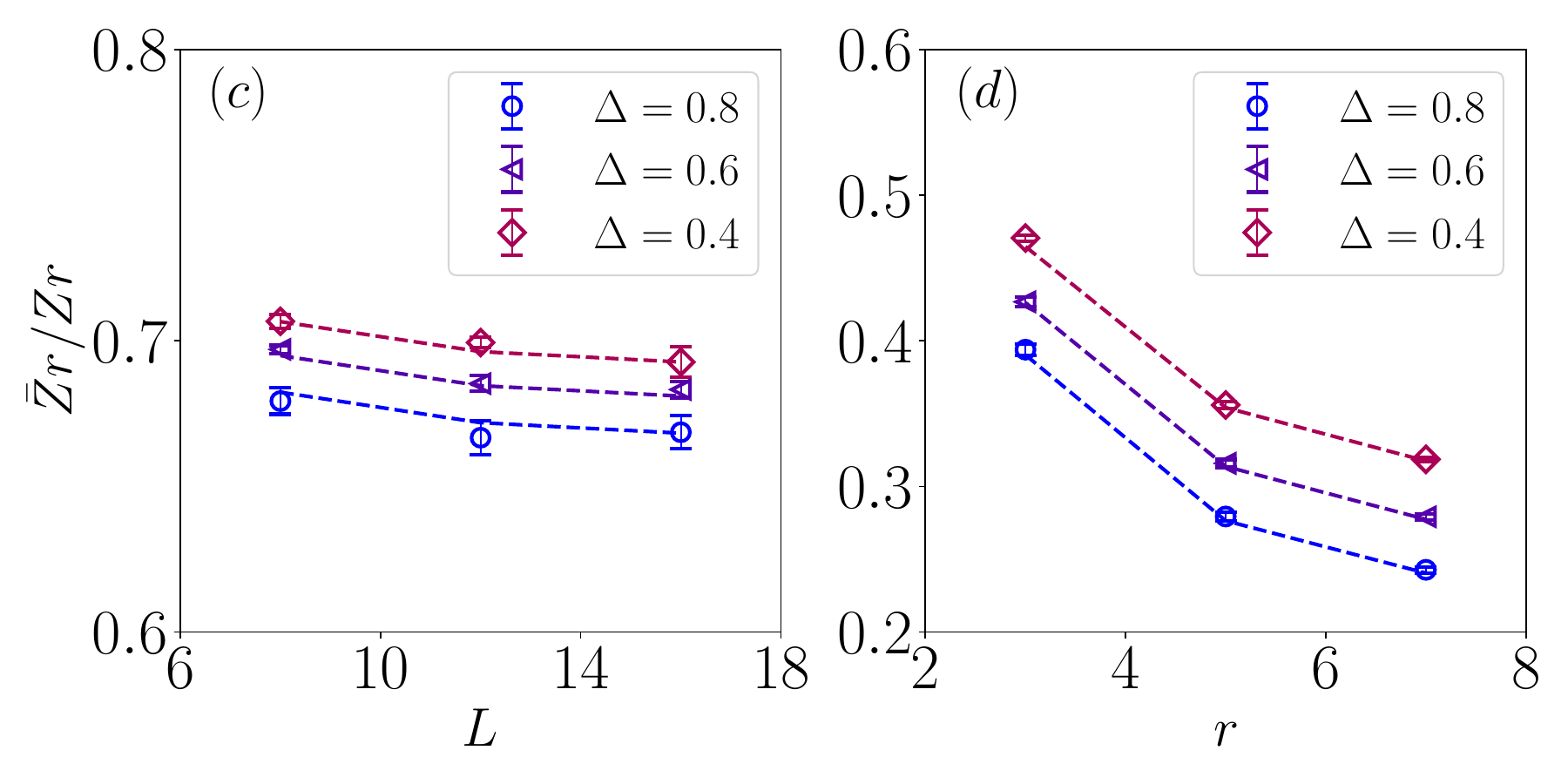}
\caption{The off-diagonal correlation measurement for $S^x$ in the 1D XXZ model comparing to the ED results (dashed line). (a) The lattice diagram for annealing along the system size $L$. We anneal the coupling $J_2$ from $0^+$ to 1. (b) The lattice diagram for annealing along the distance $r$ between $S^x_1$ and $S^x_{1+r}$. We firstly tune the coupling $J_2$ from $0^+$ to 1, then we gradually tune the coupling $J_3$ from 1 to $0^+$. (c) Annealing along the system size $L$ from 4 to 8, 12, 16 for $S^x_1S^x_2$. (d) Annealing along the distance $r$ between the $S^x_1$ and $S^x_{1+r}$ from $r=1$ to $r=3,5,7$ with fixed system size $L=16$.   Error bars denote the standard error on the mean.}
\label{fig:xxzlr}
\end{figure}

The system size $L$ and the distance $r$ can serve as annealing parameters, which are achieved via engineering the special coupling $J$ between sites and adjusting these coupling accordingly. The $\sigma^x_1\sigma^x_2$ correlation on the $L=4$ XXZ chain is regarded as a reference point or a starting seed. By incrementally adjusting the coupling $J_2$ from $0^+$ to 1,  we can effectively anneal the system size to larger size such as $L=8$, as depicted in Fig. \ref{fig:xxzlr}(a). Thus, the ratio of the PF is simply written as,
\begin{align}
\bar{Z}r&=\bigg\langle \left(\frac{J_{2}'}{J_2}\right)^{n_{J_2}}\bigg\rangle_{\bar{Z}} \nonumber\\
Zr&=\bigg\langle \left(\frac{J_{2}'}{J_2}\right)^{n_{J_2}}\bigg\rangle_{Z}
\label{eq16}
\end{align}

The reference point is $\langle S^x_1 S^x_2\rangle_{L=4}$ with open boundary condition.  As shown in Fig. \ref{fig:xxzlr}(c), we obtain the ratio of $\langle S^x_1 S^x_2\rangle$ for $L=8,12,16$ with varying $\Delta$, which is consistent with ED results. When we have obtained the $\langle S^x_1 S^x_2\rangle$ for larger systems, we employ an annealing process to vary the distance $r$ between $S^x_1$ and $S^x_{1+r}$ , in order to acquire different correlations $\langle S^x_1 S^x_{1+r}\rangle$. For instance, to calculate the $\langle S^x_1 S^x_{4}\rangle$, we construct a special XXZ chain with an extended length $L'=L+2$ (Fig. \ref{fig:xxzlr}(b)). The operator $\sigma^x_1$ and $\sigma^x_2$ are placed at site 1 and 4, respectively. Initially, we incrementally tune the coupling $J_2$ from $0^+$ to 1. The expression for ratio $\bar{Z}r_1$ and $Zr_1$ are the same as Eq.(~\ref{eq16}). Then, we reduce the coupling $J_3$ from 1 to $0^+$, effectively truncating the chain to length $L$  and removing the two terminal sites. When $J_3$ decreases, the ratio of the PF can be easily expressed as,

\begin{align}
\bar{Z}r_2&=\bigg\langle \left(\frac{J_{3}'}{J_3}\right)^{n_{J_3}}\bigg\rangle_{\bar{Z}} \nonumber\\
Zr_2&=\bigg\langle \left(\frac{J_{3}'}{J_3}\right)^{n_{J_3}}\bigg\rangle_{Z}
\label{eq17}
\end{align}

The total ratio can be considered as $\bar{Z}r/Zr=\bar{Z}r_1/Zr_1\times \bar{Z}r_2/Zr_2$. To determine the general correlation $\langle S^x_1 S^x_{1+r}\rangle$ ($r=3,5,7..$), we design a special chain with $L'=L+r-1$. The number of sites between $\sigma^x_1$ and $\sigma^x_2$ is $r-1$. Similarly, the number of sites to be removed is also $r-1$. By incrementally tuning the coupling $J_2$ and $J_3$, we can obtain the corresponding $\langle S^x_1 S^x_{1+r}\rangle$. As illustrated in Fig. \ref{fig:xxzlr} (d), we calculate the off-diagonal correlation for $L=16$. Since the $\langle S^x_1 S^x_{2}\rangle$ can be obtained via ED, we present the total ratio $\bar{Z}r/Zr=\langle S^x_1 S^x_{1+r}\rangle/\langle S^x_1 S^x_{2}\rangle$, which shows excellent agreement with the ED results.

\section*{Supplementary Note 6: Cluster Update for TFIM}
The Hamiltonian for TFIM can be written as,
\begin{equation}
H_{TFIM} =  -J\sum_{\langle i,j\rangle}\sigma^z_i \sigma^z_j -h\sum_{i}\sigma^x_i
\end{equation}

Then it can be decomposed into site and bond operators,

\begin{align}
H_{0,0}&=I  \nonumber\\
H_{-1,a}&=h(\sigma^{+}_a+\sigma^{-}_a)\nonumber\\
H_{0,a}&=h \nonumber\\
H_{1,a}&=J(\sigma^{z}_{i(a)}\sigma^{z}_{j(a)}+1)
\end{align}
where $H=-\sum^{1}_{i=-1}\sum_{a}H_{i,a}$. Here $H_{0,0}$ is the Identity operators and $i=-1,0,1$ denotes different kinds of operators: off-diagonal operator on site, diagonal operator on site and diagonal operator on bond. The subscript $a$ holds two different meaning:  for site operators $H_{0,a}$ marks the site number (for 2D lattice, $a=1,2,...,N=L^2$); and for bond operator $H_{1,a}$ index $a$ marks the bond number (for 2D lattice, $a=1,2,...,N_b=2L^2$);

According to the SSE scheme, the non-zero matrix element can be constructed via the above site operators and bond operators as follows.

\begin{align}
\langle\uparrow| H_{-1,a} |\downarrow\rangle&=\langle\downarrow| H_{-1,a} |\uparrow\rangle=h \nonumber\\
\langle\uparrow| H_{0,a} |\uparrow\rangle&=\langle\downarrow| H_{0,a} |\downarrow\rangle=h \nonumber\\
\langle\uparrow\uparrow| H_{1,a} |\uparrow\uparrow\rangle&=\langle\downarrow\downarrow| H_{1,a} |\downarrow\downarrow\rangle=2J
\end{align}
The update process contains both diagonal update and cluster update~\cite{Sandvik2003Stochastic,zhao2021Higher}. The diagonal update involves inserting or removing a diagonal operator between two states with probabilities determined by the detailed balance rules. And the cluster update is to flip all the spin and change the type of site operators on the cluster within the Swendsen-Wang scheme. During the cluster update, two key rules guide the construction of the cluster: (1) clusters are terminated on site operators $H_{-1,a}$ or $H_{0,a}$; (2) the bond operators $H_{1,a}$ belongs to one cluster. We carry out this process until all the clusters are formed. Then we flip the cluster built from the above rule with probability 1/2 (which is the Swendsen Wang cluster updating scheme).

\begin{figure}[htp]
\centering
\includegraphics[width=0.5\textwidth]{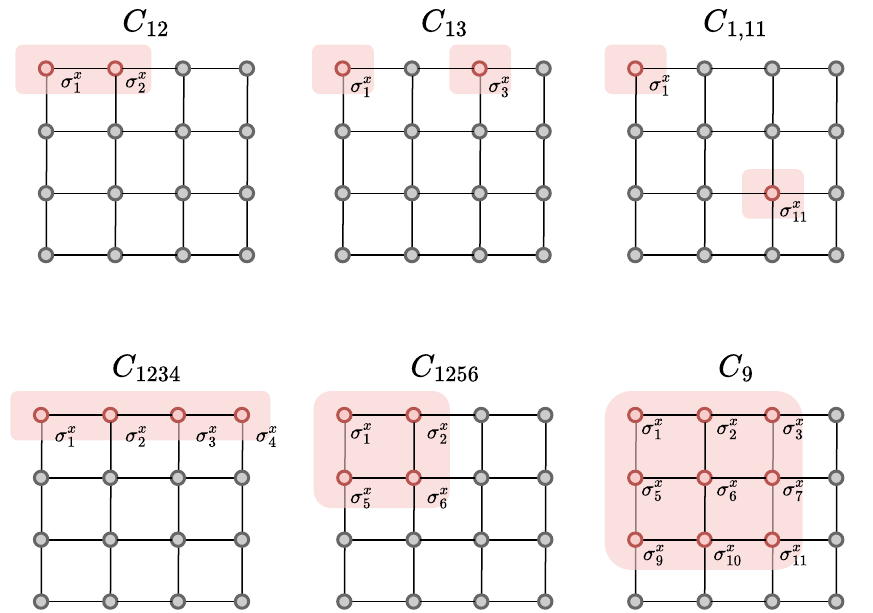}
\caption{The diagram for measuring the two-point and multi-point off-diagonal operators on the 2D TFIM. $C_{12},C_{13}, C_{1,11}$ denotes the two-point off-diagonal correlations. $C_{1234},C_{1256}, C_{9}$ denotes the multi-point correlations. $C_9$ deontes the off-diagonal correlation among spins locating at 1-3, 5-7, 9-11.}
\label{fig:ising2d}
\end{figure}

\begin{figure}[t!]
\centering
\includegraphics[width=0.5\textwidth]{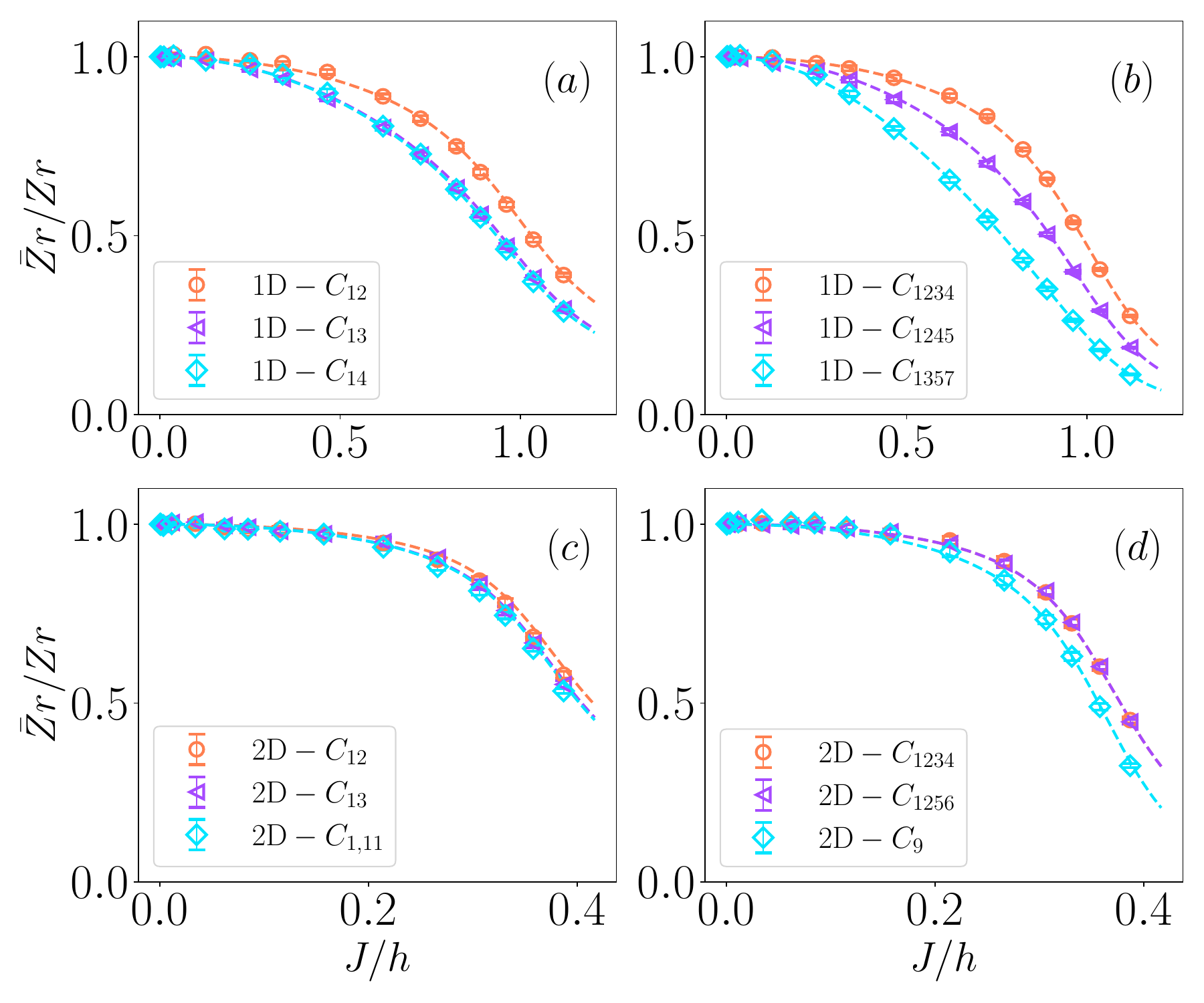}
\caption{The off-diagonal correlation measurement for $\sigma_x$ in the 1D TFIM ($L=10$ and $\beta=20$) (a-b) and 2D TFIM ($L=4$ and $\beta=32$) (c-d). All the dashed lines are the ED results. (a-b) Two-point off-diagonal correlations ($C_{ij}$) and four-point off-diagonal correlations ($C_{ijkl}$) in 1D TFIM model corresponding to the diagram of Fig.\ref{fig:1dxxzchain}. (c-d) Two-point and multi-point off-diagonal correlations in 2D TFIM model corresponding to the diagram of Fig.\ref{fig:ising2d}.  Error bars denote the standard error on the mean.}
\label{fig:ising1d2d}
\end{figure}
When inserting the measurement operator $\sigma^x$ (or many $\sigma^x$ operators) in the PF, it is necessary to insert an equal number of regular off-diagonal $\sigma^x$ operators to keep the PBC in the imaginary-time, which is similar to the XXZ model. During the diagonal update, the measurement off-diagonal operator remain unchanged, while the spins should be flipped. If we encounter the measurement operator $\sigma^x$ in the cluster update, we choose the continue straight and update the spin without changing the types of operators (as shown in Fig. \ref{fig:QMC}(b)). All other normal operators obey the standard rules of cluster update.

In the QMC simulation, we fix $h=1$ and anneal the $\bar{Z}$ and $Z$ via gradually adjusting the coupling $J$. Therefore, the ratio of the PF with and without measurement operators can be expressed as,
\begin{align}
\bar{Z}r&=\bigg\langle \left(\frac{J'}{J}\right)^{n_J}\bigg\rangle_{\bar{Z}} \nonumber\\
Zr&=\bigg\langle \left(\frac{J'}{J}\right)^{n_J}\bigg\rangle_{Z}
\end{align}

For 1D TFIM, there are two distinct phases including PM phase and FM phase, which are separated by a Ising critical point (QCP) $J/h=1$. We perform the annealing process on the coupling $J$ from $0^+$ to 1.2, spanning the critical point, to obtain the $\bar{Z}r/Zr$. The RA results indicate that the ratio $\bar{Z}r/Zr$ gradually decreases as $J/h$ increase, which is consistent with the ED results (dashed line). In the 2D TFIM, the effective QCP $J/h=0.3285$ which separates the PM phase from the FM phase. As depicted in Fig. ~\ref{fig:ising1d2d}(c) and (d), when $J/h$ gradually increases, the ratio $\bar{Z}r/Zr$ gradually decreases, which demonstrates that two point and multi-point $\sigma^x$ correlations are weakening. These observations are also in good accordance with the ED results~\cite{Phillip2017QuSpin,Phillip2019QuSpin}.

\section*{Supplementary Note 7: Error Analysis}

\begin{figure}[htp]
\centering
\includegraphics[width=0.45\textwidth]{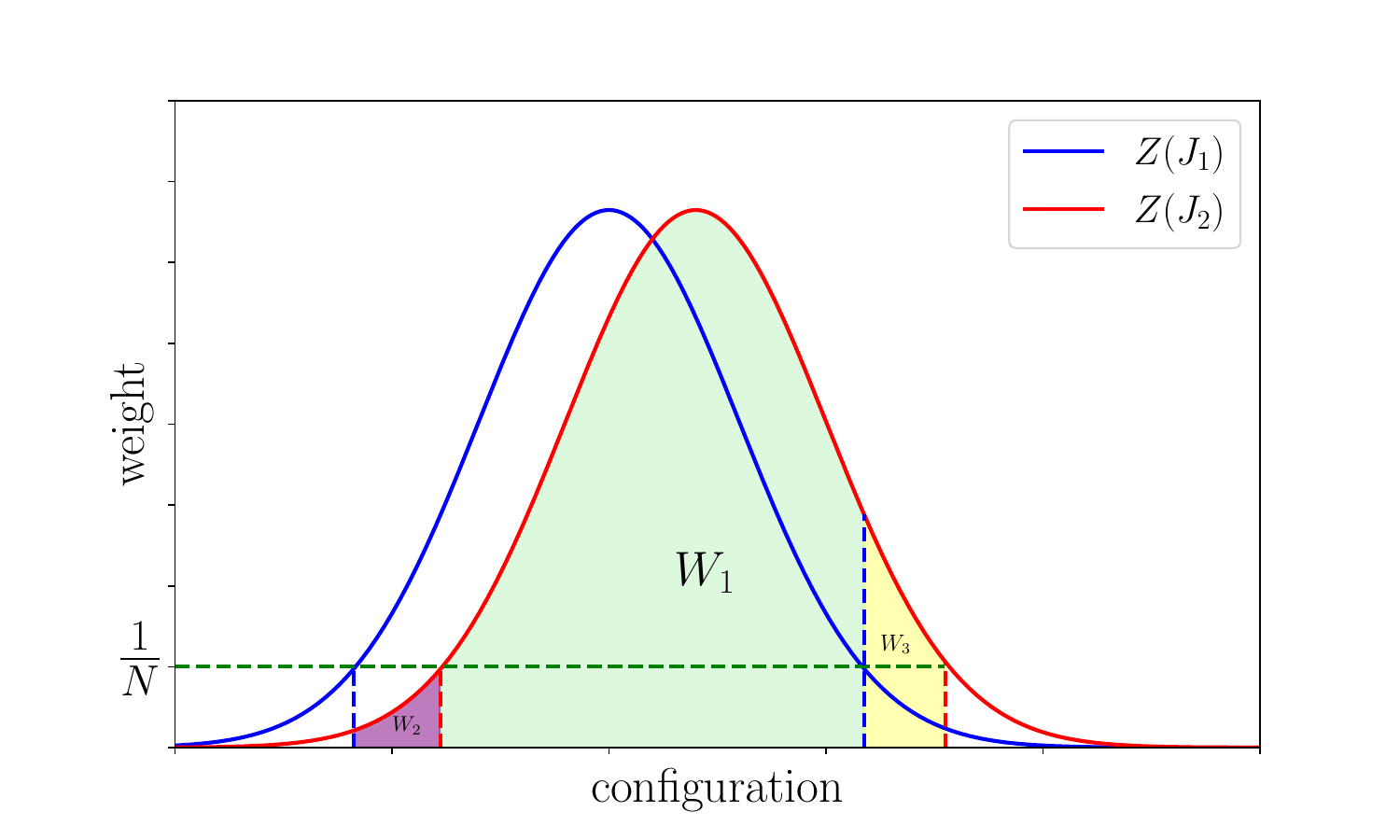}
\caption{The distribution of partition functions $Z(J_1)$ and $Z(J_2)$ with different parameter $J_1$ and $J_2$. $N$ is the number of Monte Carlo ssamplings. The region including $W_1$ and $W_2$ denotes the weights that can be sampled for $Z(J_2)$ in the reweight-annealing process. The region including $W_1$ and $W_3$ denotes the weights that can be sampled in the original directed sampling. }
\label{fig:zj1j2}
\end{figure}

In the reweight-annealing framework, the ratio between two partition functions is estimated as an expectation under the ensemble at $J_1$
\begin{align}
\frac{Z(J_2)}{Z(J_1)}=\bigg\langle \frac{W(J_2)}{W(J_1)}\bigg\rangle_{J_1}
\end{align}
The reweight-annealing estimator is unbiased when the number of Monte Carlo samples $N$ is infinite, and its accuracy can be systematically improved by increasing $N$~\cite{ding2024reweight}.

However, finite sampling introduces a systematic error when weights fall beyond the $1/N$
threshold (the region outside the sampling area corresponding to $y=1/N$ on the distribution curve in Fig.~\ref{fig:zj1j2}), making them statistically inaccessible. We set $J_1$ and $J_2$ are sufficiently close such that their ensembles overlap approximately as the two distributions (blue and red curves in Fig.~\ref{fig:zj1j2}). Configurations sampled from $P_{J_1}(\mathcal{C})$ are used to estimate properties at $J_2$. But the tail of $Z(J_2)$ distribution (region $W_3$) lies beyond reliable reach. Together with the poorly sampled region $W_2$ (beyond the area defined by $1/N$ threshold), these contribute to the systematic error ($W_2$ and $W_3$).

To control this error, we adjust the step size between $J_1$ and $J_2$ so that the ratio $\tfrac{Z(J_2)}{Z(J_1)}$ remains $\mathcal{O}(1)$, typically within $0.1$ to $10$, the details are discussed in the previous RA paper \cite{ding2024reweight}, this setting makes the computation complexity polynomial and importance sampling remained. When we estimate the distribution $Z(J_2)$ through the sampled distribution $Z(J_1)$, this ensures that the inaccessible part $W_3$ remains $\mathcal{O}(1)/N$, similarly the other over-accessible contribution $W_2$ also stay at $\mathcal{O}(1)/N$.  Therefore, the total systematic error can scale as $\mathcal{O}(1)/N \sim 1/N$. 
In contrast, the statistical error of Monte Carlo scales as $1/\sqrt{N}$. For large $N$, the systematic error is dominated by the statistical error and does not accumulate significantly even when extended over $M$ slices (the number of the incremental processes). When sampling over $M$ slices (each with an expectation value $x$), the systematic error, according to error propagation, is given by $\sqrt{M} x^{M-1}/N$ scaling with $\sim 1/N$, which remains smaller than the total statistical error $\sqrt{M} x^{M-1}/\sqrt{N}$ scaling with $\sim 1/\sqrt{N}$. Hence, the systematic error does not accumulate to a significant or uncontrollable level as $M$ increases, since it is always dominated by the statistical error. Therefore, it can be safely neglected in practice.

When tuning the parameter $J$ from weak coupling to strong coupling, the partition function will evolve rapidly near the phase transition, necessitating a greater number of slices to accurately sample the ratio. To maintain the ratio $Z(J_2)/Z(J_1)\sim O(1)$, a higher density of slices are automatically required in this regime to ensure accurate sampling of the partition function.  In particular, due to the strategy outlined in the error analysis above, the systematic error stays below the statistical error and shows no significant accumulation, even when crossing the phase transition.

To demonstrate the correction of above analysis, we show two examples below.

\section*{Supplementary Note 8: Benchmark with the worm trick method}

\begin{figure}[ht!]
\centering
\includegraphics[width=0.45\textwidth]{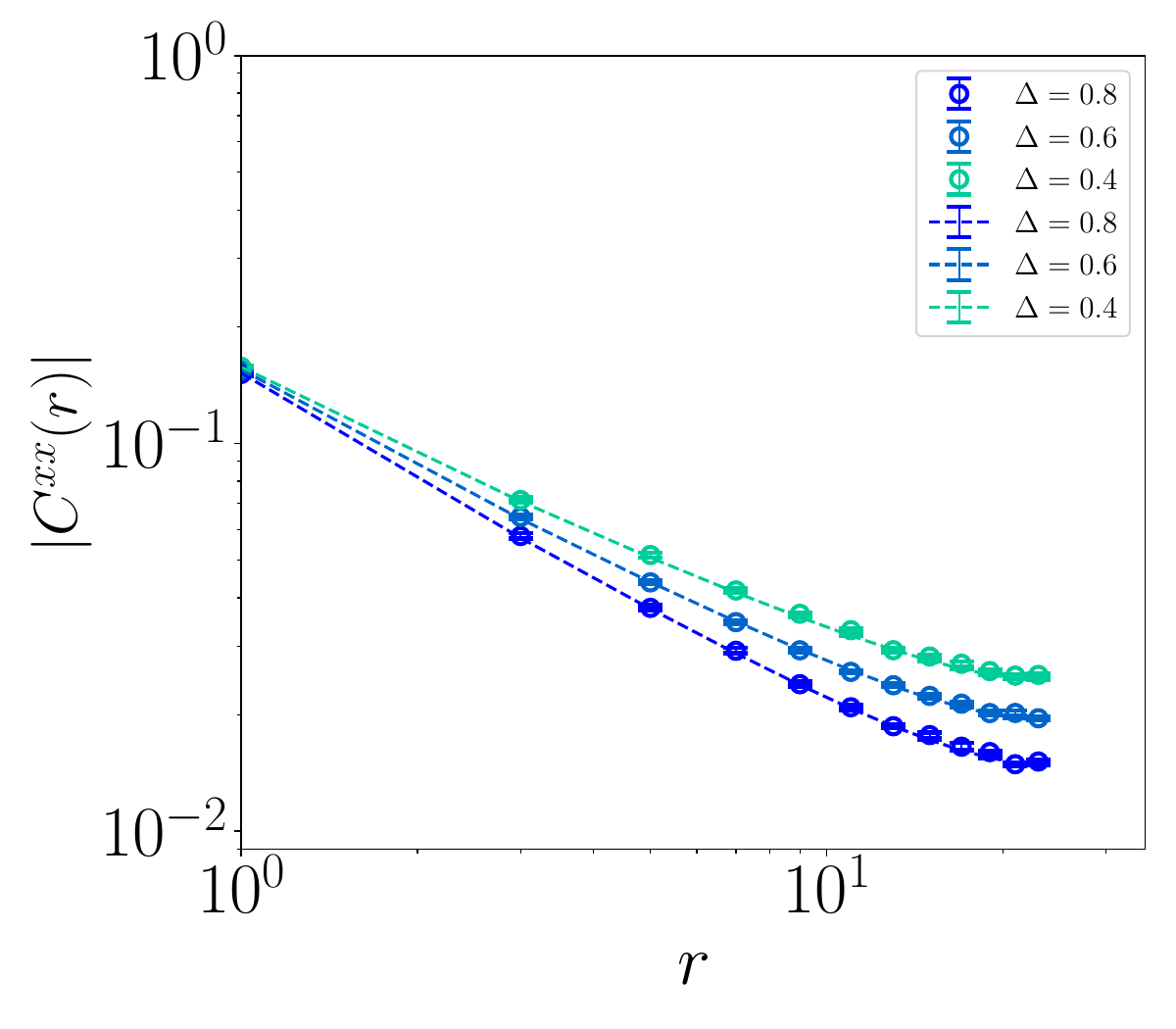}
\caption{Two point off-diagonal correlations for system size $L=48$ obtained from lattice annealing method (annealing the system size $L$ and distance $r$ between off-diagonal operators), The dashed lines represent the off-diagonal operators correlation from worm-trick method.  Error bars denote the standard error on the mean.}
\label{fig:worm}
\end{figure}

In the conventional worm-trick method, two-point off-diagonal correlation functions (Green's functions) are sampled in an extended configuration space, as detailed in Ref. \cite{wenjing2021measuring}. As shown in Fig. \ref{fig:worm}, the results for these correlation functions obtained via the lattice reweight-annealing method  (Fig.3 in main text) agree with those from the worm-trick method. This consistency demonstrates that the BRA method can be successfully extended to large system sizes while maintaining controllable error bars.

\section*{Supplementary Note 9: Benchmark with the disorder operators in the 2D TFIM}

\begin{figure}[ht!]
\centering
\includegraphics[width=0.5\textwidth]{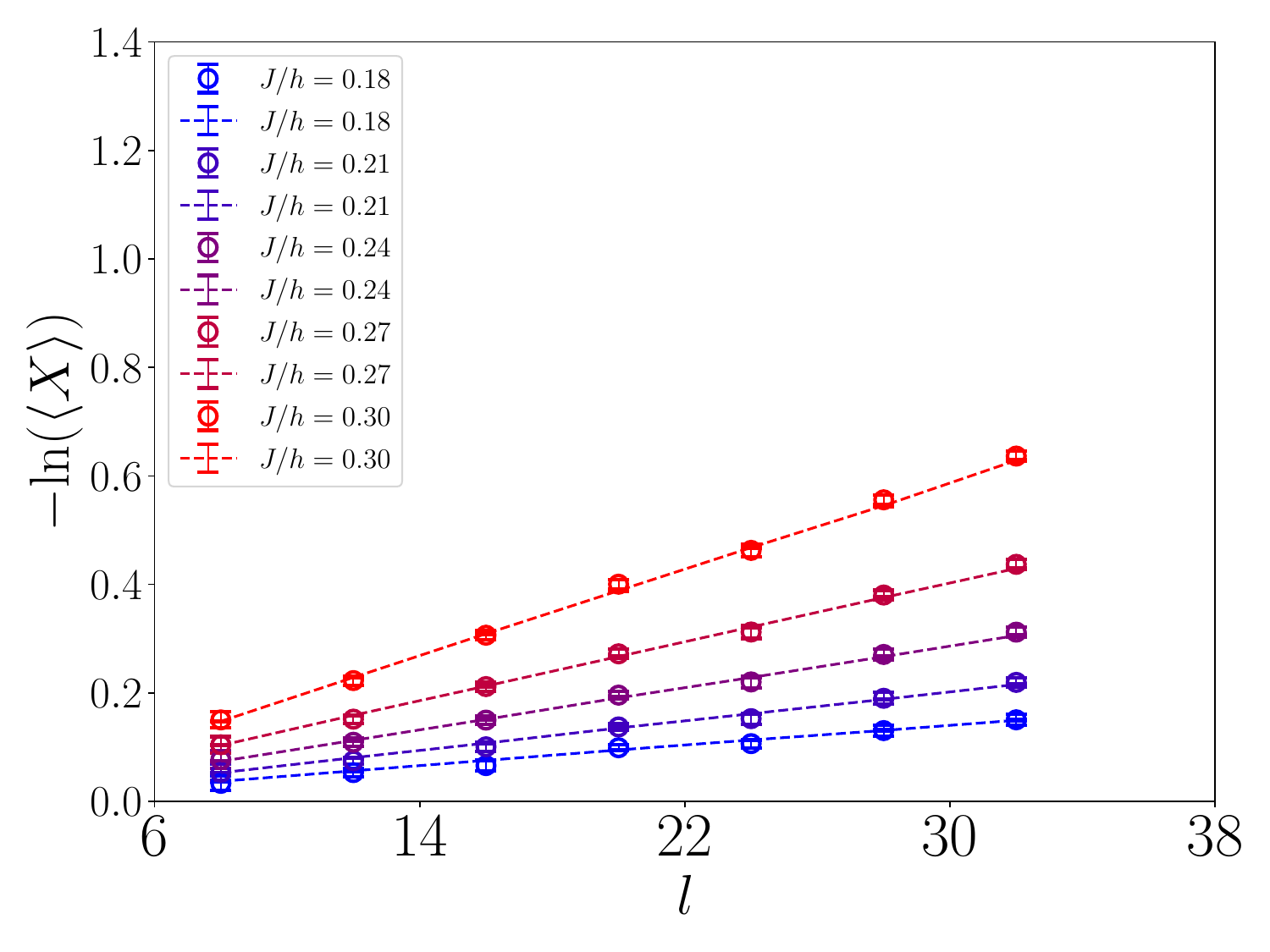}
\caption{$Z_2$ disorder operator $\langle X \rangle =\langle \prod_{i\in M} \sigma^x_i \rangle$ obtained from the reweight annealing method (circle) and directed measurement in the $\sigma_x$ basis (dashed line) in the 2D TFIM model.  Error bars denote the standard error on the mean.}
\label{fig:disop}
\end{figure}

In the 2D transverse-field Ising model (TFIM), the $Z_2$ disorder operator $\langle X \rangle = \langle \prod_{i \in M} \sigma^x_i \rangle$ is diagonal in the $\sigma^x$ basis and can thus be measured directly. The Hamiltonian can be simulated efficiently in this basis using the directed-loop algorithm, as detailed in Ref.~\cite{zhao2021Higher}. As shown in Fig.~\ref{fig:disop}, the disorder operator values obtained via the reweighting-annealing method agree closely with those from direct measurement, with all results falling within error bars. This agreement confirms that the systematic error does not accumulate significantly in large systems.


\end{document}